# The application of sub-seasonal to seasonal (S2S) predictions for hydropower forecasting

Robert M. Graham[1], Jethro Browell[1,3], Douglas Bertram[2], Christopher J. White[2]

1. Department of Electronic and Electrical Engineering, University of Strathclyde, Glasgow
2. Department of Civil and Environmental Engineering, University of Strathclyde, Glasgow
3. School of Mathematics and Statistics, University of Glasgow

**Key points:**

- S2S probabilistic inflow forecasts for hydropower reservoir in Scotland demonstrate skill up to 6-weeks ahead
- Skilful S2S probabilistic inflow forecasts may be derived from S2S ensemble predictions without an additional hydrological model
- Stylised cost-model demonstrates economic value of S2S probabilistic inflow forecasts through improved water management

**Funding Statement**

This research was funded through the University of Strathclyde's Low Carbon Power and Energy program, which is a strategic industrial collaboration between the University, Scottish and Southern Energy, Scottish Power and Wood. Jethro Browell is supported by EPSRC Innovation Fellowship (EP/R023484/1). We thank Andrew F. Low and Richard Hearnden from SSE Renewables for their valuable input on this project and providing the historical inflow record for the case study reservoir. The authors declare no conflicts of interest.

**Data availability**

The 2019 hindcast dataset for the ECMWF extended-range forecasts was accessed from the S2S prediction project (http://s2sprediction.net/) database hosted by the ECMWF (https://apps.ecmwf.int/datasets/data/s2s/levtype=sfc/type=cf/). ERA5 records were accessed from the NCAR-UCAR Research Data Archive (https://rda.ucar.edu/).The NAO index data were accessed from the Climate Research Unit at the University of East Anglia (https://crudata.uea.ac.uk/cru/data/nao/). Forecast skill scores were calculated using the package 'SpecsVerification' in R. The normalised inflow record for the case study reservoir will be published together with this manuscript.

**Abstract**

Inflow forecasts play an essential role in the management of hydropower reservoirs. Forecasts help operators schedule power generation in advance to maximise economic value, mitigate downstream flood risk, and meet environmental requirements. The horizon of operational inflow forecasts is often limited in range to ~2-weeks ahead, marking the predictability barrier of deterministic weather forecasts. Reliable inflow forecasts in the sub-seasonal to seasonal (S2S) range would allow operators to take proactive action to mitigate risks of adverse weather conditions, thereby improving water management and increasing revenue. This study outlines a method of deriving skilful S2S inflow forecasts using a case study reservoir in the Scottish Highlands. We generate ensemble inflow forecasts by training a linear regression model for the observed inflow onto S2S ensemble precipitation predictions from the European Centre for Medium-range Weather Forecasting (ECMWF). Subsequently, post-processing techniques from Ensemble Model Output Statistics are applied to derive calibrated S2S probabilistic inflow forecasts, without the application of a separate hydrological model. We find the S2S probabilistic inflow forecasts hold skill relative to climatological forecasts up to 6-weeks ahead. The inflow forecasts hold greater skill during winter compared with summer. The forecasts, however, struggle to predict high summer inflows, even at short lead-times. The potential for the S2S probabilistic inflow forecasts to improve water management and deliver increased economic value is confirmed using a stylised cost model. While applied to hydropower forecasting, the results and methods presented here are relevant to broader fields of water management and S2S forecasting applications.

## 1. Introduction

Hydropower reservoirs provide a crucial source of flexibility to many national power systems, and are likely to become increasingly important in the future as the capacity of non-dispatchable electricity sources, such as wind and solar, continue to increase (Ho et al. 2020; Huertas-Hernando et al. 2017). Inflow forecasts play an integral role in the management and operations of hydropower resources (Yang et al. 2017; Ahmad and Hossain 2019; Bazile et al. 2017; Magnusson et al. 2020; Huertas-Hernando et al. 2017). Hydropower plant operators require reliable inflow forecasts to schedule power generation in advance. Inflow forecasts allow operators to maximise power generation and the value of water in a reservoir by preventing spillage and optimising generation to periods of peak energy price (Ahmad and Hossain 2019; Ødegård et al. 2019; Huertas-Hernando et al. 2017). In addition, operators manage water levels and discharge from reservoirs to mitigate downstream flood risks and meet environmental requirements, such as minimum flow rates (Ahmad and Hossain 2019; Yang et al. 2017; Huertas-Hernando et al. 2017). Operations are also constrained by technical limitations, including plant capacity and maintenance requirements.

Extending the horizon of operational inflow forecasts offers potential to improve planning and water management decisions (Magnusson et al. 2020; Alexander et al. 2021; Bazile et al. 2017; Yang et al. 2017; Contreras et al. 2020). Hydropower operators in Scotland currently use deterministic inflow forecasts, which are derived from deterministic precipitation forecasts and cover periods up to 14-days ahead. Reliable forecast information covering longer horizons would allow plant operators to take early, proactive measures to mitigate risks (Sene 2016; Vitart et al. 2017; Ahmad and Hossain 2019; Klemm and McPherson 2017). For example, the actions of a hydropower plant operator could differ widely dependent on the information provided by a 7-day forecast showing below average inflow rates, and a 28-day forecast indicating increased probability of above average inflows. These actions may ultimately lead to very different economic outcomes, depending on the accuracy of the forecast information.

Precipitation and streamflow forecasts are essential for effective water management and are utilised by a wide variety of users, beyond hydropower applications (Sene 2016; Baker et al. 2019). For example, forecasts play an important role in the management of public water supplies and the activation of early warning and response systems for floods and droughts (Arnal et al. 2018; Svensson et al. 2015; Bell et al. 2017; White et al. 2017; Vitart et al. 2017). Precipitation forecasts are also used by farmers determine the optimal time for planting and harvesting, and provide information on the availability of water for irrigation (Klemm and McPherson 2017; White et al. 2017). As with inflow forecasts for hydropower applications, extending the horizon of precipitation and streamflow forecasts could improve water management, by allowing users of these forecasts to take proactive rather than reactive measures to mitigate risks (Vitart et al. 2017).

For many river systems globally, detailed hydrological observations of the catchment area can be used to produce skilful streamflow forecasts on seasonal or even annual timescales (Wood and Lettenmaier 2008; Harrigan et al. 2018; Svensson et al. 2015; Svensson 2016; Bell et al. 2017). Inherent forecasting skill for these river systems originates from a large water storage capacity within the catchment area that acts as a long-term memory. This water storage capacity may include features such as mountain snow cover, aquifers, and soil moisture. For example, recent studies have explored how snow depth observations could be used to improve hydropower forecasts and operations in Norway (Magnusson et al. 2020; Ødegård et al. 2019). However, the steep hillsides, low soil water capacity and milder climate of the Scottish Highlands contribute to a number of fast flowing river systems. Here, skilful streamflow forecasts are dependent foremost on accurate meteorological forecasts, rather than the initial hydrological conditions of the catchment area (Harrigan et al. 2018; Svensson et al. 2015; Svensson 2016; Bell et al. 2017).

The horizon of deterministic weather forecasts for the mid-latitudes is limited to the medium-range, covering periods up to 10 to 14-days ahead (Zhang et al. 2019). Beyond this range, deterministic forecasts hold no skill due to the chaotic nature of the climate system (Branković et al. 1990; Buizza and Leutbecher 2015; Zhang et al. 2019). Nonetheless, the field of Sub-seasonal to Seasonal (S2S) forecasting, which covers horizons between 2-weeks and 2-months ahead, is rapidly emerging (White et al. 2017; Vitart et al. 2017; Vitart and Robertson 2018; Vitart 2014; Merryfield et al. 2020). Reliable forecast information in the S2S range has been identified as high value to a wide range of industries and users (White et al. 2017). S2S forecasts are notoriously challenging, because the time-scale is sufficiently long that the climate system's memory of the atmospheric initial conditions is lost, yet too short for ocean variability to play an important role (Vitart 2004, 2014; Vitart et al. 2017). Beyond the S2S range, is the more established field of seasonal forecasting that covers horizons from 3 to 6 months ahead (Palmer and Anderson 1994). Skill for seasonal forecasts is derived primarily from ocean variability (Merryfield et al. 2020).

Over recent decades, there has been extensive research and development to improve the reliability and accuracy of S2S forecasts (Branković et al. 1990; Vitart et al. 2017; Vitart 2014; Merryfield et al. 2020). Key sources of predictability in the S2S range have been identified as the Madden–Julian oscillation (MJO) in the tropics and stratosphere-troposphere coupling (Merryfield et al. 2020; Vitart et al. 2017; Vitart 2014; Vitart and Robertson 2018; Lee et al. 2019; Woolnough 2019; Orsolini et al. 2011). Advances in ensemble prediction systems have been central to the success of S2S forecasting (Vitart 2004). Ensemble prediction systems produce multiple predictions of the future weather, or ensemble members, which account for uncertainties in the initial atmospheric conditions and model physics. Rather than a single prediction of the future weather, ensemble prediction systems provide information on the probability of different weather patterns emerging. In recent years, there has been a well documented increase in skill of S2S forecasts, which has been attributed to advances in ensemble prediction systems and the representation of parameterised processes in Numerical Weather Prediction models (Vitart 2014). The majority of the World Meteorological Organisations Global Producing Centres now produce operational S2S forecasts (Vitart et al. 2017).

In this study, we use S2S ensemble predictions to develop a S2S forecasting system for hydropower reservoirs and focus on a case study in Scotland. Section 2 describes the data used in this study, including the case study inflow record and S2S ensemble predictions. Section 3 documents the methods used to generate and evaluate the inflow forecasts. The results are presented in Section 4, which evaluates the skill of the S2S inflow forecasts. Section 5 discusses potential sources of skill for the S2S inflow forecasts, and the economic value of these forecasts for the hydropower sector. This is followed by a summary of the main conclusions in Section 6.

## 2. Data

### 2.1 Observed reservoir inflow

The case study site is a large hydropower reservoir in northwest Scotland (Figure 1). The reservoir sits at an elevation of 200 m and is surrounded by steep hillsides, without trees. The observational record for the case study reservoir covers the period from 2009 to 2019. The dataset includes hourly time-series of the reservoir water level and power generated from the facility. From these data, we calculate a historical observed inflow record.

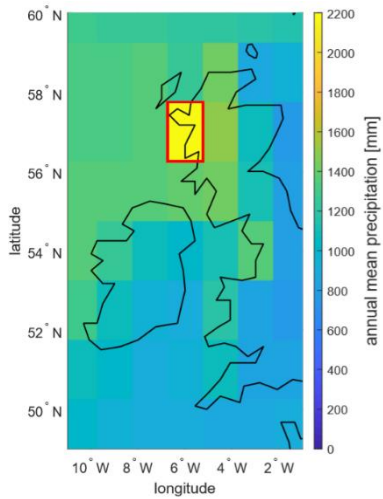

*Figure 1 mean annual precipitation for days 1-7 of S2S ensemble precipitation forecasts shown on 150 km grid. Red box indicates grid cell of the case study reservoir.*

The water level and power generation data were cleaned by removing data points outside the physical limits of these fields and spurious spikes in the time-series. Specifically, data were flagged where the data value or temporal derivative exceeded threshold values. Approximately 5% of the hourly data points were removed during this cleaning process.

Based on the power generation time-series, discharge (m³/s) from the reservoir was calculated as follows:

$$discharge\ (t) = \frac{power\ (t)}{efficiency\ (t) \times density \times gravity \times height\ (t)}$$

where *power* is the power in Watts (kg m²/s³) at time *t*; *density* is the density of water (1000 kg/m³); and *gravity* is 9.81 m/s². The *efficiency* is the turbine efficiency, and *height* is the net head (m). The turbine efficiency and net head vary in time, and are dependent on both the reservoir water level and operating power. Values for the turbine efficiency and net head were estimated at each time interval based on the observed power output and reservoir water level, using efficiency curves provided by the hydropower operator. In addition to water discharged used to generate power, compensation flows are released from a separate outlet to sustain minimum flow rates in the downstream river system. These flows are required by legislation to support fish, other wild life and vegetation in the catchment.

The volume of water in the reservoir was calculated from the reservoir water level record, using a storage curve equation for the facility. We differentiated the volume of water with respect to time, to calculate the rate of change of water volume per second (units m³/s). Finally, the *net inflow* (m³/s) was calculated by adding together the discharge and rate of change of volume, in addition to documented compensation flows from the reservoir:

$$net\ inflow\ (t) = discharge\ (t) + volume\ change\ (t) + compensation\ flow\ (t)$$

The hourly time-series of net inflow contains a lot of noise. However, a clear signal emerges for the daily mean inflow (Figure 2). For consistency, inflow is always presented as a mean inflow rate (m³/s) averaged over the forecast horizon. To respect anonymity of the site, values presented here are normalised by the annual mean inflow rate for the facility. Negative values of net inflow exist in the observed inflow record. These indicate a net reduction of water in the reservoir (excluding discharge from the facility), through processes such as evaporation or infiltration combined with low precipitation.

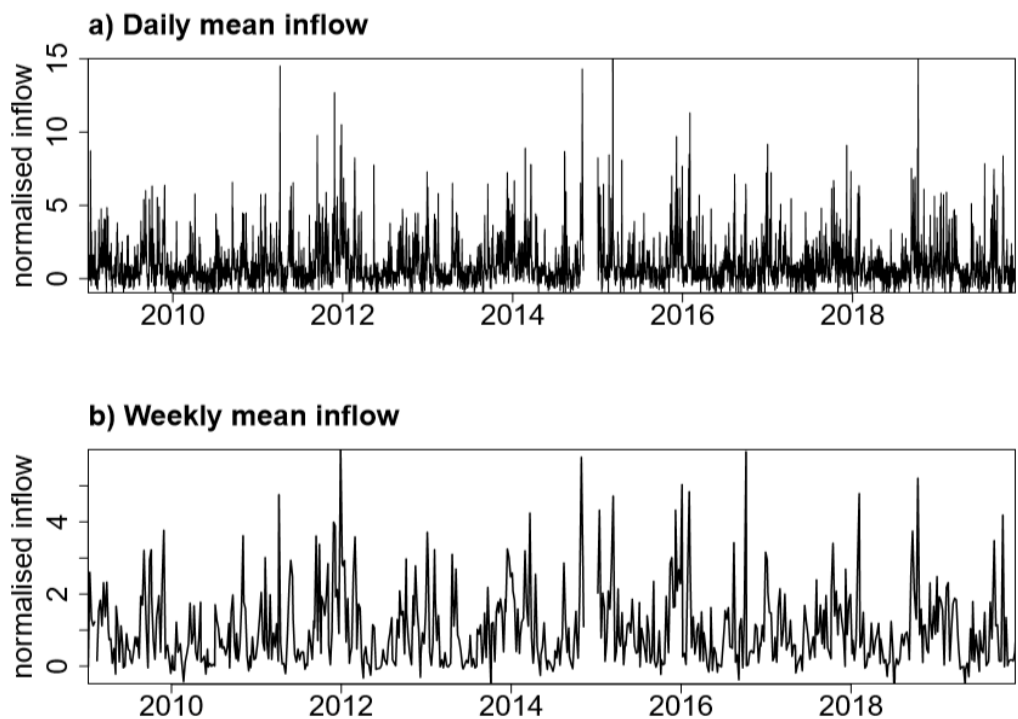

*Figure 2 Time-series of (a) daily and (b) weekly mean net inflow rate for the case study reservoir. Values are normalised by the annual mean inflow rate.*

## 2.2 S2S ensemble weather predictions

We generate and evaluate historical S2S inflow forecasts using the 2019 hindcast dataset for the European Centre for Medium-range Weather Forecasting (ECMWF) extended-range forecasts (Vitart 2014). This dataset includes historical S2S weather predictions for the period 1999-2018. These ensemble predictions are issued twice per week, on Mondays and Thursdays, as an extension to the medium-range (10-day) predictions. The predictions have a forecast horizon of 46-days (6.5 weeks); a temporal resolution of 6 hours; and horizontal resolution of 36 km (Vitart 2014). The historical predictions are comprised of 11 ensemble members, which are produced by perturbing the initial atmospheric conditions and model physics (i.e. the parameters of unresolved processes in the model). The spread of ensemble members provides information on uncertainty of the prediction and likelihood of different weather patterns emerging.

The hindcast dataset was accessed from the Sub-seasonal-to-Seasonal Prediction Project database (Vitart et al. 2017). S2S ensemble prediction datasets from eleven forecasting centres are available through this data portal. While the ECMWF extended-range predictions have a model resolution of 36 km, all datasets stored in the S2S prediction project archive are interpolated onto a common grid with a horizontal resolution of approximately 150 km (Figure 1). This spatial averaging is justified because the primary source of predictability in S2S forecasts originates from largescale weather patterns. Nonetheless, the coarser resolution of these files compared with the original model resolution will negatively impact our assessment of skill for the raw and uncalibrated precipitation predictions and benchmark inflow forecasts for the case study reservoir. This is due to smoothing of localised orographic rainfall patterns in the mountainous region of the case study reservoir. However, the application of the coarser resolution dataset is likely to have minimal impact on the final calibrated probabilistic inflow forecasts.

*Table 1: Forecast horizons for the S2S inflow forecasts and S2S ensemble predictions evaluated in this study.*

| Forecast Name | Forecast horizon |
|---|---|
| Forecast Week 1 | Days 1 – 7 |
| Forecast Week 2 | Days 8 – 14 |
| Forecast Week 3 | Days 15 – 21 |
| Forecast Week 4 | Days 22 – 28 |
| Forecast Week 5 | Days 29 – 35 |
| Forecast Week 6 | Days 36 – 42 |
| 2 Week Forecast | Days 1 – 14 |
| 3 Week Forecast | Days 1 – 21 |
| 4 Week Forecast | Days 1 – 28 |
| 5 Week Forecast | Days 1 – 35 |
| 6 Week Forecast | Days 1 – 42 |

S2S ensemble predictions of largescale and convective precipitation were extracted from the 150 km grid at the location of the case study reservoir, using bilinear interpolation (Figure 1). The variable we use is the total precipitation, which is calculated by summing the convective precipitation and largescale precipitation. For each ensemble member, we average the predicted precipitation rate over the horizon for the different S2S inflow forecasts (Table 1).

## 2.3 Atmospheric reanalysis

We use atmospheric reanalysis data to evaluate the skill of the S2S ensemble weather predictions and explore the observed relationship between precipitation and inflow at the case study reservoir. The atmospheric reanalysis used is ERA5, from the ECMWF (Hersbach et al. 2020). ERA5 has a horizontal resolution of approximately 30 km, which is comparable to the model resolution of the S2S ensemble predictions (Hersbach et al. 2020; Vitart 2014).

We extract the total precipitation (convective precipitation + largescale precipitation) from ERA5 at the location of the case study reservoir using bilinear interpolation. These hourly data are averaged into daily mean precipitation rates for comparison with the observed inflow dataset.

In this study, we assume that the atmospheric reanalysis dataset represents the true precipitation record from the case study site. Atmospheric reanalyses represent the largescale variability of the weather and climate well. However, spatial and temporal biases exist for many variables, particularly in regions of complex terrain due to coarse spatial resolution and inability to fully resolve local orographic effects (Lorenz and Kunstmann 2012).

# 3. Methods

## 3.1 Deriving inflow forecasts from S2S ensemble weather predictions

We derive inflow forecasts by training a linear regression model for the observed inflow onto the S2S ensemble predictions of precipitation. The linear regression model is fitted to a training dataset and verified on "out-of-sample" data. This ensures that error metrics are representative of the true predictive performance and not a result of over-fitting (Messner et al. 2020). We employ a cross-validation scheme to produce and evaluate as many out-of-sample forecasts as possible. Out-of-sample forecasts are produced for each year where observations are available, using a model trained on all years excluding both the forecast year and subsequent year. For example, when evaluating inflow forecasts for 2015, inflow observations and S2S ensemble weather predictions from 2015 and 2016 are excluded from the training dataset. This system avoids possible artificial enhancement of the forecast skill through long-term memory in the climate system.

For each horizon of the S2S inflow forecasts, a linear regression model for inflow is trained on the average precipitation rates from week 1 (days 1-7) of the S2S ensemble weather predictions and corresponding 7-day mean observed inflow rate. Uncertainty in S2S ensemble weather predictions increases with lead-time, as reflected by a growing spread of the ensemble members. If the linear regression model for inflow is trained on mean precipitation rates from later horizons of the S2S ensemble weather predictions, as opposed to the first week, the relationship between predicted precipitation and observed inflow is weakened. As a result, all inflow forecasts generated from the regression model approach climatological values.

We generate forecasts of the average inflow rate for 11 different forecast horizons that range in length from 1 to 6 weeks (Table 1). For each ensemble member of the S2S ensemble weather predictions, the predicted precipitation rate is temporally averaged over the horizon of the inflow forecast. The average precipitation rate from this ensemble member is then applied to the linear regression model. Through this method, we use the linear regression model to generate an ensemble of S2S inflow forecasts for the 11 different forecast horizons. We refer to these as the benchmark ensemble inflow forecasts.

To improve on the benchmark S2S ensemble inflow forecasts, we apply post-processing techniques to produce calibrated probabilistic S2S inflow forecasts that better capture seasonal trends and non-linear relationships. We employ Ensemble Model Output Statistics (EMOS), also known as semi-parametric regression (Gneiting and Katzfuss 2014). Our method builds on that outlined by Scheuerer (2014) to post-process precipitation forecasts for Germany. The model of Scheuerer (2014) is an example of a Generalized Additive Model for Location, Scale, and Shape (GAMLSS), which is an extremely flexible modelling framework in which the parameters of a given parametric distribution are modelled as linear (generalised) additive models (Rigby and Stasinopoulos 2005).

The parameters of a chosen probability distribution for the inflow forecasts are modelled as functions of features derived from the $K$ ensemble members $m_{1,t}, \ldots, m_{K,t}$ at time $t$ of the benchmark inflow forecast. These features are the ensemble mean inflow

$$\bar{m}_t = \frac{1}{K} \sum_{k=1}^{K} m_{k,t}, \tag{1}$$

the fraction of ensemble members less than or equal to zero

$$\overline{\mathbf{1}_{\{m=0\}}}_t = \frac{1}{K} \sum_{k=1}^{K} \mathbf{1}_{\{m_{k,t} <= 0\}}, \tag{2}$$

and the ensemble mean difference (Scheuerer, 2014)

$$\overline{\Delta m}_t = \frac{1}{K^2} \sum_{k=1}^{K} \sum_{k'=1}^{K} |m_{k,t} - m_{k',t}|. \tag{3}$$

The ensemble mean difference is preferable to other measures of dispersion, such as standard deviation and interquartile range, because it is based on absolute rather than squared differences and is sensitive to all ensemble members (Scheuerer 2014).

Here, we choose the Zero Adjusted Gamma distribution $f_{\text{ZAGA}}(\cdot)$ of inflow $y$ is effectively a mixture of a conventional Gamma distribution and a probability mass at zero. It is given by

$$f_{\text{ZAGA}}(y|\mu, \sigma, \nu) = \nu \delta(y) + (1-\nu) \left[ \frac{1}{(\sigma^2 \mu)^{1/\sigma^2}} \frac{y^{\frac{1}{\sigma^2}-1} e^{-y/(\sigma^2 \mu)}}{\Gamma(1/\sigma^2)} \right] \tag{4}$$

where $\mu$ and $\sigma$ are the shape and scale parameters of the conventional Gamma distribution, respectively, and $\nu$ is the probability of $y = 0$. $\delta(\cdot)$ is the Dirac delta function, which is equal to 1 when $y = 0$ and zero otherwise, and $\Gamma(\cdot)$ is the Gamma function. We model the parameters of the Zero Adjusted Gamma distribution as

$$\log(\mu_t) = \beta_{1,0} + \beta_{1,1} \bar{m}_t + \beta_{1,2} \overline{\mathbf{1}_{\{m=0\}}}_t + s_1(t), \tag{5}$$

$$\log(\sigma_t) = \beta_{2,0} + \beta_{2,1} \bar{m}_t + \beta_{2,2} \overline{\Delta m}_t + s_2(t), \text{ and} \tag{6}$$

$$\text{logit}(\nu_t) = \beta_{3,0} + \beta_{2,1}\bar{m}_t, \tag{7}$$

where $s_i(t)$ are cyclic cubic regression splines with a period of one year that model smooth seasonal variation, and $\beta_{i,j}$ are parameters to be estimated. This post-processing was performed using the *gamlss* package in R (Rigby and Stasinopoulos 2005). The choice of distribution and additive model structure was guided by expert judgement and cross-validation on training data.

In contrast with the linear regression model, a separate EMOS model is used for each forecast horizon to account for the evolving properties of the ensemble forecast with increasing lead-time. As described above, the forecast probability distribution for inflow is modelled as a Zero Adjusted Gamma distribution and the parameters of this distribution are dependent on the three input statistics (ensemble mean inflow, fraction of forecasts equal to zero, and ensemble mean difference) derived from the benchmark ensemble inflow forecasts, as well as a smoothed function for the day of the year. Zero Adjusted Gamma distributions do not allow negative values. However, in practice, negative values of inflow exist due to the effects of evaporation and infiltration during prolonged periods without precipitation. We therefore add an offset all observed inflow values, which corresponds to the largest negative inflow rate within each training period. This offset is then subtracted from the modelled probability distributions. The final S2S probabilistic inflow forecasts, following the EMOS post-processing, provide the probability of the average inflow rate over the forecast horizon falling above or below any given threshold.

## 3.2 Evaluating the skill of S2S inflow forecasts and ensemble weather predictions

The skill of the S2S ensemble weather predictions and probabilistic inflow forecasts is evaluated using the fair Continuous Ranked Probability Skill Score (fCRPSS) (Manrique-Suñén et al. 2020; Ferro 2014). This is a common method used for evaluating ensemble and probabilistic forecasts (Buizza and Leutbecher, 2015; Harrigan et al., 2018; Scheuerer, 2014). The Continuous Ranked Probability Score is similar to the Mean Absolute Error for deterministic forecasts and measures the difference between the forecast and observed cumulative density functions. The CRPSS is a measure of the average Continuous Ranked Probability Score over the forecast evaluation period, benchmarked against the use of climatological forecasts. The fair version of this skill score (fCRPSS) is designed to reward ensembles that behave as though they are sampled from the same distribution as the verifying observations, which is particularly important for small ensemble sizes (Ferro 2014; Manrique-Suñén et al. 2020).

An fCRPSS of 1 indicates a perfect forecast, while a score of 0 or below indicates no skill relative to the use of climatological forecasts. Between these extremes, we classify the skill scores (fCRPSS) less than 0.15 as fair, between 0.15 and 0.30 as good, and greater than 0.30 as very good (fCRPSS > 0.30). This follows the convention used by the S2S4E project (https://s2s4e.eu/). These thresholds are somewhat arbitrary as the value of a forecasts will depend on the use-case under consideration, which motivates the proceeding section of this paper.

The climatological forecasts that we use as a benchmark to evaluate the S2S ensemble weather predictions and inflow forecasts correspond to the climatological distribution of precipitation and inflow rates for the month of the forecast, averaged over the duration of the forecast horizon. Consistent with the method used for splitting observations into training and evaluation datasets, we exclude the observations from the forecast year and subsequent year from the climatological distributions. For example, climatological inflow forecasts for January 2015 would include all historic observations from January except years 2015 and 2016.

### 3.3 Evaluating economic value of inflow forecasts

In addition to evaluating the statistical skill of the S2S probabilistic inflow forecasts, we assess the potential economic value of these forecasts through improved water management. To measure forecast value, we develop a stylised cost model based on the classical *"News Vendor"* optimisation problem (Khouja 1999). The cost model is guided by current operational practice, with the underlying principle of maintaining the reservoir at a target water level for the time of year.

In the stylised cost model framework, costs fall into two categories. The first category includes costs incurred from precautionary actions taken to maintain the reservoir at its target water level, based on forecast information. The second category includes costs incurred due to the actual observed inflow, following any precautionary actions taken. An example of category 1 costs would be if an inflow forecast indicates an increased probability of high inflows. Based on this forecast information, operators extend the planned generation schedule by five hours. This increases discharge from the reservoir, in order to accommodate the forecast inflow. However, the additional power generation takes place during periods of off-peak energy prices, requiring power to be sold at a lower unit rate, and therefore represent an opportunity cost. Category 2 costs relate to any deviations from the target water level, at the end of each forecast period. For example, if the actual observed inflow is greater than the discharge from the reservoir, despite any precautionary adjustments to the generation schedule, the water level in the reservoir will rise. This reduces the potential for impounding future inflows. To prevent future spillages, operators decide to increase discharge from the reservoir. However, this requires extending power generation into off-peak periods, resulting in a lower unit rate of power sold and thus an opportunity cost.

The optimal (risk-neutral) water management decision for each forecast period is to take the precautionary action (i.e. adjust the power generation schedule) that minimises the expected costs based on the information provided by the inflow forecasts. If the inflow forecasts are reliable and skilful, this will deliver reduced costs relative to using climatological forecasts. The cost model is applied independently to each inflow forecast in the case study dataset (i.e. the 11 forecast horizons, with two forecasts per week from 2009-2018). The relative economic value of three types of forecasts (climatological, deterministic and probabilistic), are evaluated by comparing the total costs incurred for all forecasts over the complete time-series. Deterministic inflow forecasts are defined here as the median (p50) of the predictive distribution of the S2S probabilistic inflow forecasts. Climatological forecasts are also deterministic, following current operational practice, and are defined here as the median (p50) of the historical inflow observations.

In this model framework, the economic value of the different forecasts is sensitive to the choice of peak and off-peak energy price. To explore this dependency, the cost model is run for a range of price differentials (peak – off-peak) from £5/MWh to £100/MWh, while the peak price is kept constant at £50/MWh. The same differential may also result from a higher peak price and non-negative off-peak price. Therefore, the relative change in water value with price differential is more relevant than the absolute value. To test the significance of the results, we apply a bootstrap method to resample the 10-year time-series 1000 times (Messner et al. 2020). All results shown indicate the spread of 2 standard errors based on bootstrap resampling, which approximately represents the 90% confidence interval. For a complete description of the cost model set-up, we refer to the supplementary material.

## 4. Results

### 4.1 Relationship between precipitation and observed inflow

Based on a cross correlation between the ERA5 daily mean precipitation rate and observed inflow, we find a peak correlation coefficient of 0.76 with a 0-day lag-time (Figure 3a). The 0-day lag indicates that most precipitation within the catchment area takes less than a day to enter the case study reservoir. Thus this is a highly responsive catchment area, which is consistent with the steep terrain surrounding the reservoir and relatively low soil water capacity in western Scotland (Bell et al., 2017; Harrigan et al., 2018; Svensson, 2016; Svensson et al., 2015). The cross correlations between the reanalysis precipitation and observed inflow are asymmetric, with higher correlation coefficients for negative lag times (Figure 3). This demonstrates that the observed inflow is more strongly correlated with past precipitation, as opposed to future, which is to be expected. The correlations between precipitation and observed inflow are higher if we consider weekly average precipitation and inflow rates, giving a peak correlation coefficient of 0.90 (Figure 3b). Several physical processes can explain the higher correlation coefficients found for longer averaging periods, including delays between precipitation and inflow due to freezing temperatures (e.g. snowfall and frozen soil water) and changing soil water content. In addition, temporal averaging will smooth out errors in the reanalysis precipitation record.

The high correlation coefficients found here for the ERA5 precipitation rate and observed inflow demonstrate a strong physical coupling between precipitation and inflow at the case study reservoir. This justifies the choice of linear regression model to forecast inflow. These results also indicate that ERA5 provides a reliable precipitation record for the case study site.

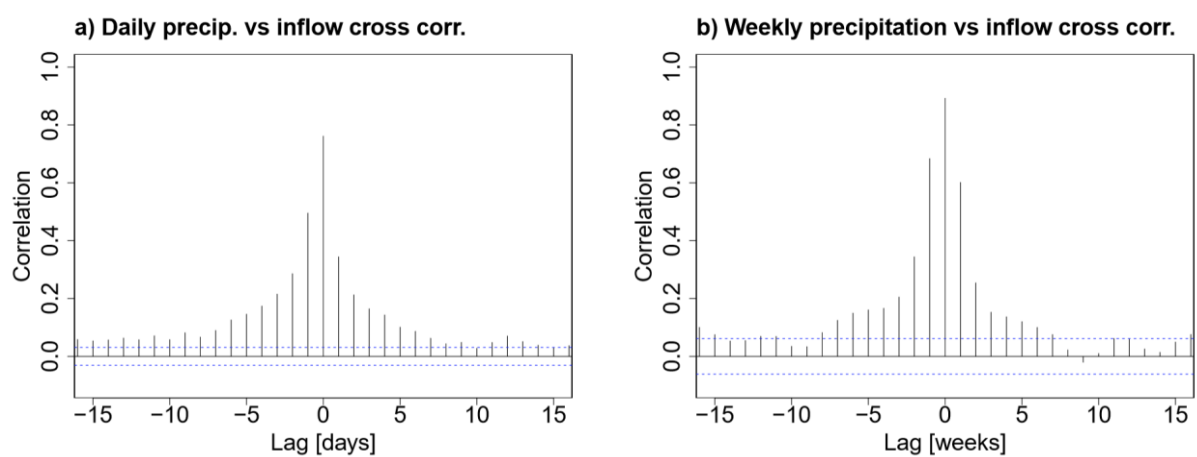

*Figure 3* Cross correlation of observed inflow and ERA5 precipitation at the case study reservoir. a) Daily mean, and b) weekly mean precipitation and inflow rates. Negative lags indicate how past precipitation is correlated with present inflow.

### 4.2 Skill of S2S ensemble weather predictions

We calculate an fCRPSS of 0.45 for the average precipitation rate in week 1 (days 1-7) of the S2S ensemble weather predictions, benchmarked against the ERA5 precipitation record at the case study reservoir (Figure 4a). This demonstrates very good skill. However, the skill of the ensemble predictions decreases rapidly with increasing lead-time. We find an fCRPSS of 0.10 for the average precipitation rate in week 2 (days 8-14), reflecting fair skill (Figure 4a). Moving beyond week 2 and into the S2S range, skill scores are negative (Figure 4a). This means that the climatological distributions are more reliable than the S2S ensemble predictions.

The timing of precipitation events is a major source of uncertainty in S2S ensemble weather predictions. If we consider predictions of the average precipitation rate over periods of time longer than one week, this source of uncertainty is reduced and greater skill is available (Figure 4). Some users may be more interested predictions of total precipitation over an entire month, rather than the total precipitation within each week of that month. Significant forecast skill is available to these users. For example, predictions of the average precipitation over a 4-week period (days 1-28) hold comparable skill to forecasts for the 7-day average precipitation rate during week 2 (days 8-14) (Figure 4). Fair skill (fCRPSS=0.04) even exists for predictions of the average precipitation over the coming 6-weeks (days 1-42) (Figure 4). It is important to note that the primary source of skill for these extended average precipitation predictions originates from weeks one and two of the S2S ensemble weather predictions. The skill and reliability of precipitation totals for weeks 1-2 of the S2S ensemble predictions provides a strong indication whether the period as whole will be wetter or drier than average.

### 4.3 Skill of S2S inflow forecasts

Consistent with the ensemble weather predictions, we identify skill for the 7-day average benchmark ensemble inflow forecasts for the horizons of week 1 and week 2, but no further (Figure 4c). The benchmark inflow forecasts hold good skill (fCRPSS=0.25) for week 1 (days 1-7) of the forecast, and fair skill for week 2 (Figure 4c). For longer averaging periods, the benchmark S2S ensemble inflow forecasts also demonstrate fair skill (fCRPSS of 0.02) relative to climatological distributions for six-week (days 1-42) average inflow forecasts (Figure 4d).

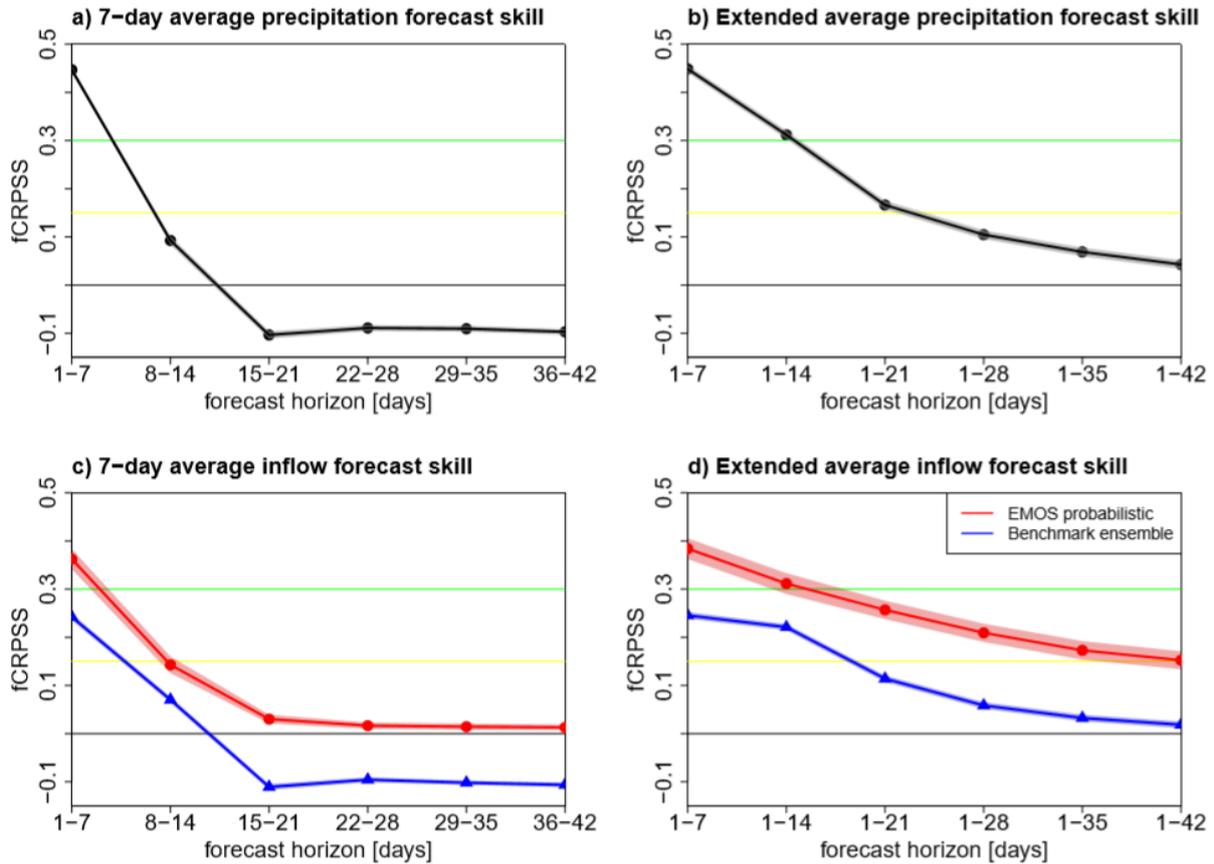

*Figure 4 Skill of the precipitation and inflow forecasts at the case study reservoir, evaluated for different forecast horizons (Table 1). a) Skill of the S2S ensemble weather predictions for 7-day average precipitation rates of different lead times. b) Skill of the S2S ensemble weather predictions for extended average precipitation rates from 1 to 6 weeks ahead. C) Skill of the S2S benchmark ensemble (blue) and EMOS probabilistic (red) inflow forecasts for 7-day average inflow rates of different lead times, corresponding to Forecast Weeks 1 – 6. b) Skill of the S2S inflow forecasts for extended average inflow rates from 1 to 6 weeks ahead. Forecast skill is measured using the fair Continuous Ranked Probability Skill Score, using the SpecsVerification package in R. Shaded areas indicate +/- 2 standard deviations. Skill of the precipitation predictions is evaluated over the 1999-2018 hindcast period. Inflow forecast skill is evaluated over the observation record from 2009-2018. An fCRPSS value > 0 is classed as fair forecast skill, progressing to good forecast skill for values above 0.15 (yellow line) and very good for values above 0.3 (green line).*

Weeks 1 and 2 of the benchmark ensemble inflow forecasts largely capture the observed variability of inflow at the case study reservoir (Figure 5). However, the ensemble inflow forecasts are under-dispersive. When comparing the observed inflow and the spread of ensemble members for the benchmark inflow forecasts, it is clear that the observed inflow frequently falls close to or outside the limits of the ensemble spread (Figure 5b). This is particularly common during periods when the observed inflow is low. When comparing the distribution of the observed inflow and forecast inflow, it is clear that the benchmark inflow forecasts under-predict the occurrence of both high and low inflow events (Figure 6).

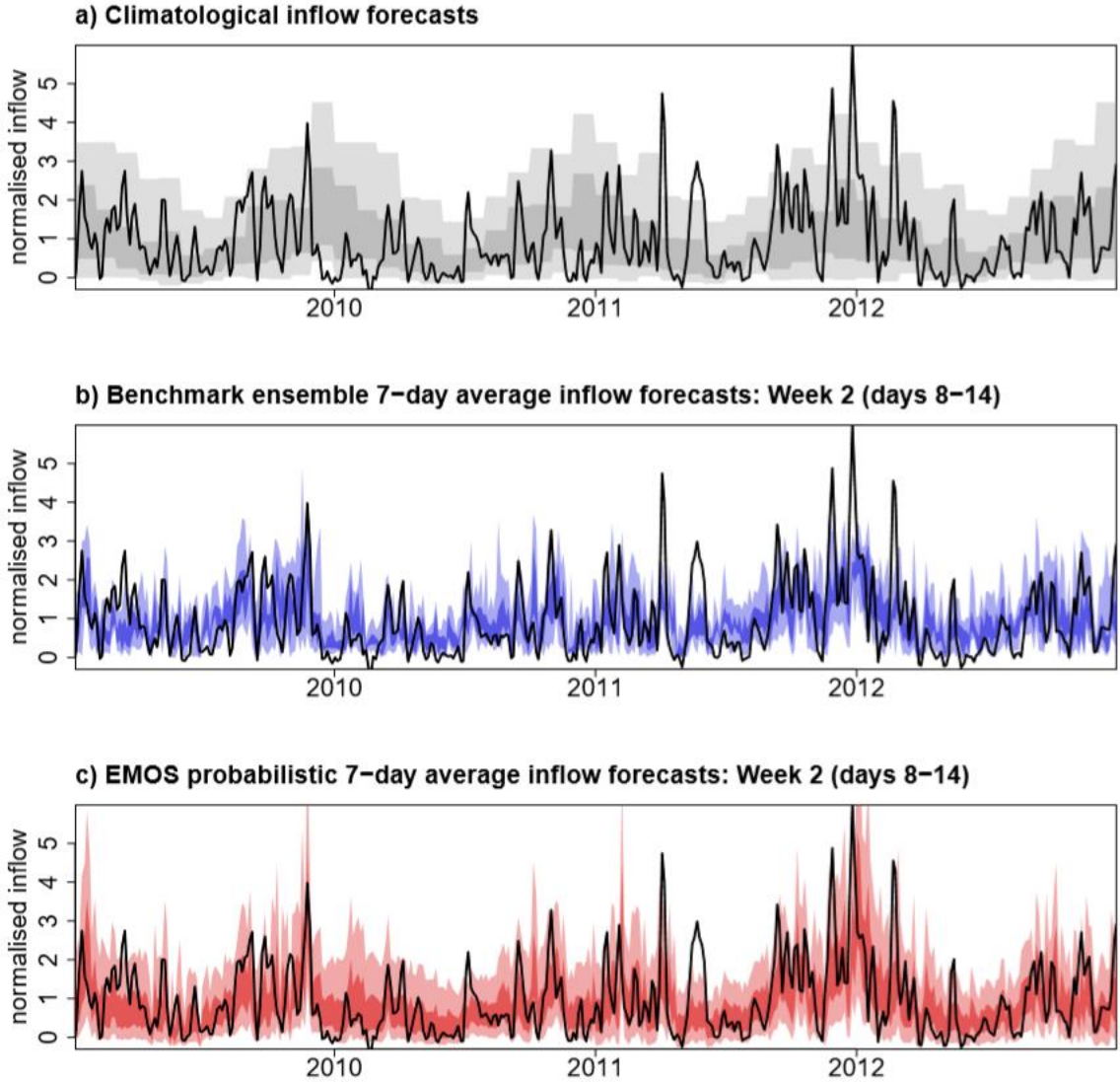

*Figure 5: Time-series of observed 7-day mean inflow rate for the case study reservoir. Inflow values are normalised by the annual mean inflow a) Climatological inflow forecasts (based on historical inflow observations). b) Benchmark ensemble inflow forecasts for Week 2 (days 8-14). c) EMOS probabilistic forecasts for Week 2. Shaded regions indicate the 50% and 90% prediction intervals.*

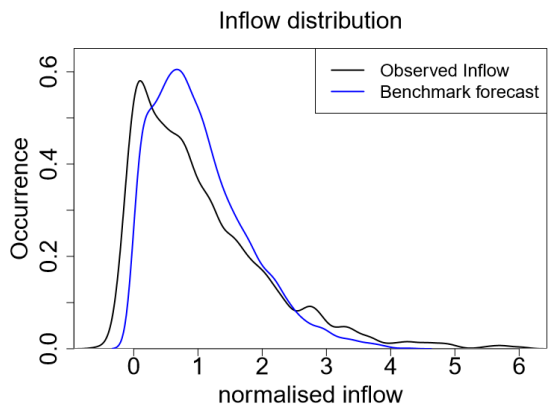

*Figure 6: Comparison of the observed distribution of weekly mean inflow (black) and Week 1 (days 1-7) of the benchmark ensemble inflow forecasts (blue) over the 2009-2018 observation period. Inflow values are normalised by the annual mean inflow rate.*

The EMOS post-processing calibrates the inflow forecasts, and thus correct the under-dispersion of the benchmark S2S ensemble inflow forecasts. Hence, the observed inflow more frequently falls within the forecast distribution of the probabilistic inflow forecasts (Figure 5b-c). Post-processing also improves the probabilistic accuracy of the S2S inflow forecasts. Probabilistic accuracy means, for example, that the empirical proportion of observations falling below the 10% quantile of the forecast distribution should be 10%. Reliability diagrams show that the S2S probabilistic inflow forecasts perform well in this measure, with limited deterioration in probabilistic accuracy for increasing lead-time (Figure 7). Importantly, post-processing cannot correct errors that originate from the S2S ensemble weather predictions failing to predict an event, such as the high inflows during summer 2011 (Figure 8b-c).

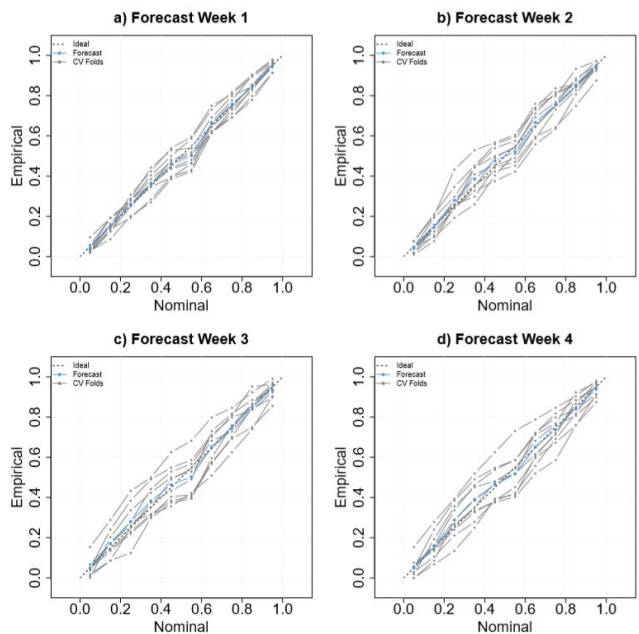

*Figure 7: Reliability diagrams for weeks 1 to 4 of the EMOS probabilistic inflow forecasts.*

The EMOS post-processing greatly increases the skill of the benchmark inflow forecasts (Figure 4c-d). The fCRPSS for the S2S probabilistic inflow forecasts is significantly higher than the benchmark ensemble forecasts for all forecast horizons. The largest increase in skill score for the 7-day average inflow forecasts is during Week 3 (days 15-21), where the fCRPSS increases by 0.14. For Forecast Weeks 4 – 6 the CRPSS increases by approximately 0.1. Importantly, 7-day average S2S probabilistic inflow forecasts hold fair skill relative to climatology for Week 3 to Week 6, with fCRPSS values between 0.01-0.03 (Figure 4a). Hence, EMOS post-processing breaks the forecasting skill horizon of 14 days lead-time. Nonetheless, while significant, the skill scores for forecasts week 3 to week 6 are relatively low and suggest only a marginal improvement upon the use of climatological inflow distributions.

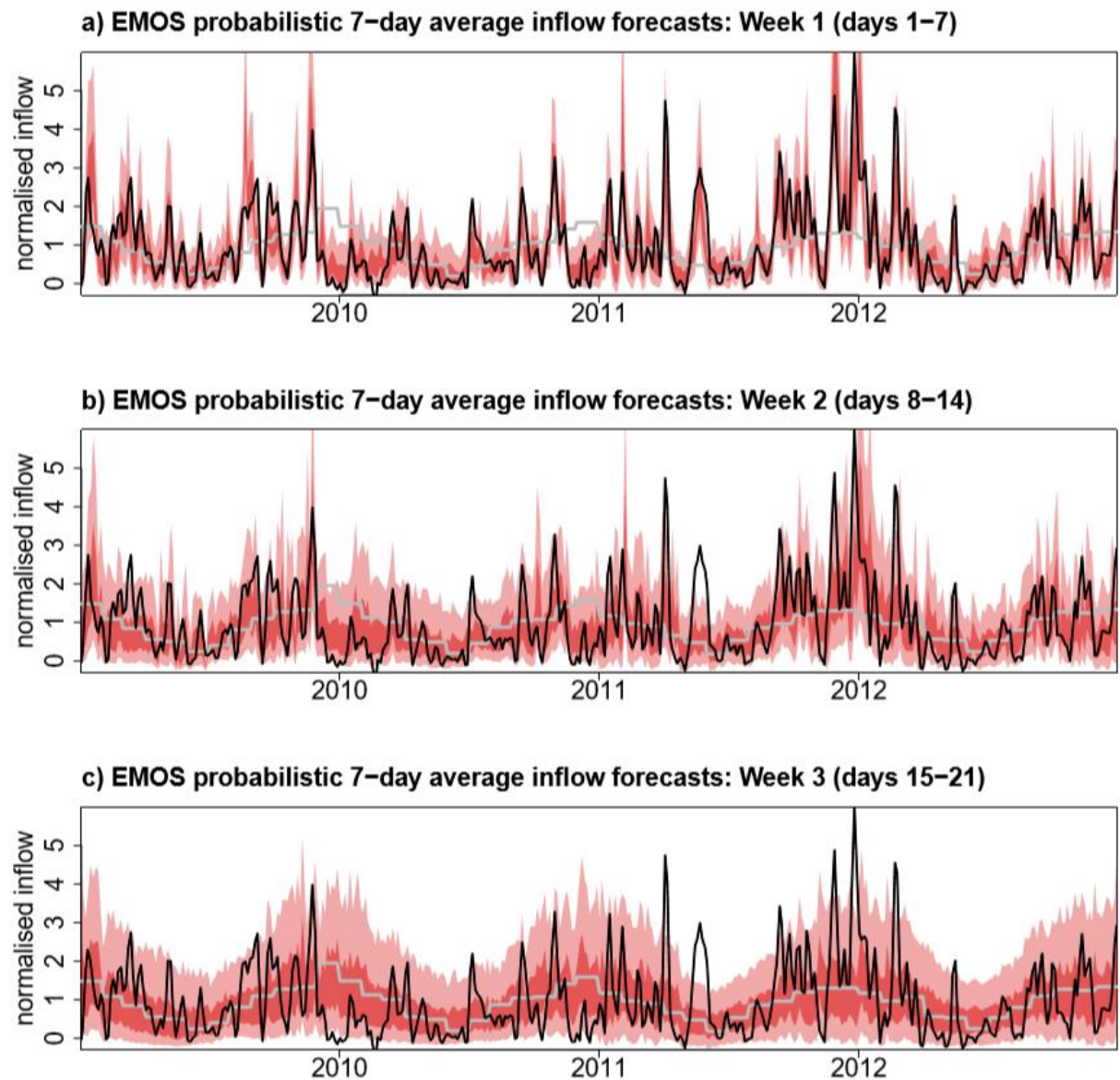

*Figure 8: Time-series of observed weekly mean inflow rate (black) and EMOS probabilistic inflow forecasts for a) Week 1 (days 1-7), b) Week 2 (days 8-14), and c) Week 3 (days 15-21). Shaded areas indicate the 50% and 90% prediction intervals. Grey line shows the deterministic climatological forecast. Inflow values are normalised by the annual mean inflow. Full 2009-2018 observation period is shown in supplementary material.*

Significantly greater skill is available for users interested in forecasts of the average inflow rate over periods longer than 7-days. Notably, 6-week average probabilistic inflow forecasts (days 1-42) hold good skill relative to climatological forecasts, and are comparable in skill to the 7-day average probabilistic forecasts for Week 2 (days 8-14) (Figure 4d and Figure 9). The skill of long-range average inflow forecasts decreases gradually with increasing forecast horizon, compared with a rapid reduction in forecast skill with increasing lead-time for weekly average inflow forecasts (Figure 4).

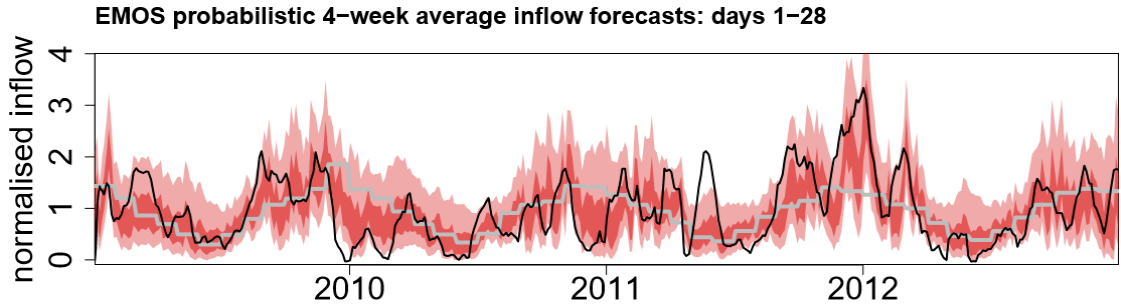

*Figure 9 Time-series of observed 28-day mean inflow at the case study reservoir (black line), and EMOS probabilistic 4-week (days 1-28) average inflow forecasts (red). Shaded areas indicate the 50% and 90% prediction intervals. Grey line shows the deterministic climatological forecast. Inflow values are normalised by the annual mean inflow. Full 2009-2018 observation period is shown in supplementary material.*

## 4.4 Impact of atmospheric and seasonal variability on forecast skill

The seasonal variability of weather patterns and modes of atmospheric variability can affect forecast skill. Here, we evaluate the skill of the 7-day average S2S probabilistic inflow forecasts separately for the extended summer (April – September) and winter (October – March) seasons. We find that the inflow forecasts consistently hold higher skill during winter than summer, across all lead-times (Figure 10a). However, the difference in skill is not always significant. The largest difference in skill scores between the summer and winter seasons is during forecast week 2 (days 8-14). Here, winter time forecasts hold good skill, while summer forecasts demonstrate fair skill (Figure 10a). Skill scores for weeks 3-6 of the summer forecasts are not significantly greater than zero. Similar patterns are found for the benchmark inflow forecasts and S2S ensemble weather predictions (not shown).

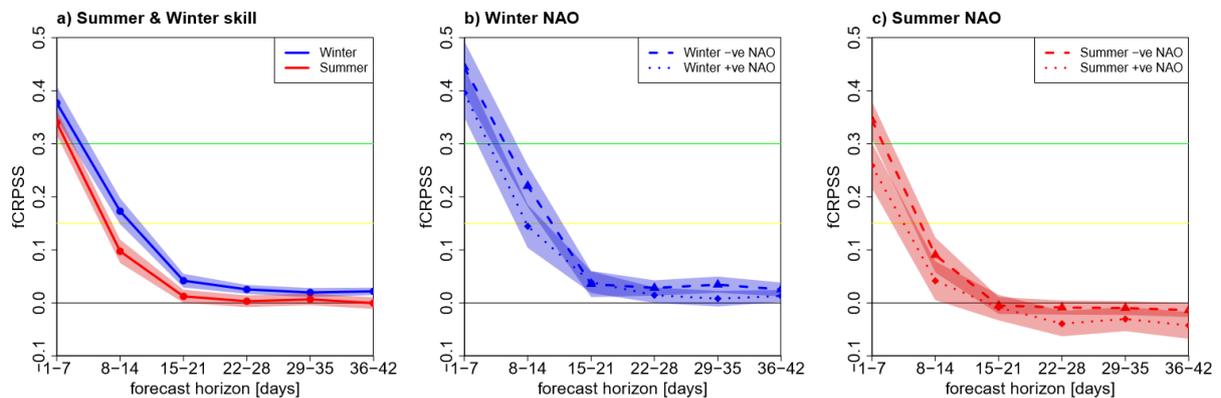

*Figure 10: Impact of seasonal and atmospheric variability on the skill of S2S probabilistic inflow forecasts for different forecast horizons (Table 1). a) Forecast skill for summer (April-September, red) and Winter (October –March, blue). b) Impact of North Atlantic Oscillation on winter inflow forecast skill. c) Impact of North Atlantic Oscillation on summer inflow forecast skill. Positive NAO periods correspond to an NAO index > +0.4 (dotted lines), and a negative NAO correspond to an NAO index < -0.4 (dashed lines). Forecast skill is measured using the fair Continuous Ranked Probability Skill Score. Forecast skill is assessed over the 2009-2018 observation period. Shaded areas indicate +/- 2 standard deviations. Forecast skill is measured using the fair Continuous Ranked Probability Skill Score, using the SpecsVerification package in R. An fCRPSS value > 0 is classed as fair forecast skill, progressing to good forecast skill for values above 0.15 (yellow line) and very good for values above 0.3 (green line).*

We further evaluate the skill of the S2S probabilistic inflow forecasts for different phases of the North Atlantic Oscillation (NAO). It is important to note that this published NAO index (Jones et al. 1997) is unknown at the time the inflow forecasts are issued. The NAO index is a standardised measure of the surface pressure difference between the Icelandic low and the Azores high, west of Portugal (Wanner et al. 2001). We calculate the forecast skill for each phase of the NAO by considering forecasts for which the observed monthly average NAO index during the forecast period is greater than +0.4 (NAO positive) or less than -0.4 (NAO negative).

The highest skill scores for the 7-day average S2S probabilistic inflow forecasts are during winter months with a negative NAO index (Figure 10). However, differences between the forecast skill during positive and negative phases of the NAO during winter are typically not significant. The largest difference in skill score for the different phases of the NAO during winter are for the forecast horizon of week 2 (days 8-14), when forecasts during negative phases of the NAO hold good skill and forecasts during positive phases hold fair skill (Figure 10b). The summer inflow forecasts also have higher skill scores during negative phases of the NAO compared with positive phases (Figure 10c). However, for both phases of the NAO during summer, inflow forecasts hold no skill beyond week two.

These analyses of how the phase of the NAO impacts the skill of inflow forecasts must be treated with caution, due to the relatively short observational time-series. Thus, there is a limited number of positive and negative NAO events for each season. This is reflected by the large standard deviation for the fCRPSS values (Figure 10).

## 5. Discussion

### 5.1 The use and value of S2S probabilistic inflow forecasts for the hydropower sector

There is a growing understanding within the S2S forecasting community that statistical skill may not translate to a good measure of forecast value for end-users (Brayshaw et al. 2020; Ødegård et al. 2019; Anghileri et al. 2019). For a forecast to deliver value, the forecast information must be sufficiently reliable to consistently improve operational outcomes over an extended period of time. In the present application for the hydropower sector, the S2S probabilistic inflow forecasts must be sufficiently reliable to allow hydropower operators to optimise power generation to peak price periods, and therefore increase the average unit rate achieved for power generated.

In the stylised cost model framework, the value of the S2S probabilistic inflow forecasts follow similar patterns to those found for the forecast skill. The mean value of the 7-day average S2S probabilistic inflow forecasts decreases with increasing lead-time (Figure 11a). The 7-day average probabilistic inflow forecasts for Week 2 (days 8-14) deliver additional value of £2.20 / MWh, relative to the use of climatological forecasts, when averaged over the full range of price differentials simulated. This value decreases to £1.40 / MWh for the Week 6 (days 36-42) forecasts. When applying forecasts of longer duration, the 2-week mean (days 1-14) probabilistic inflow forecasts deliver an average additional value of £2.70 / MWh, decreasing to £1.10 / MWh for the 6-week (days 1-42) average forecasts (Figure 11b). Overall, we find that the S2S probabilistic inflow forecasts consistently deliver additional economic value, relative to the use of deterministic climatological forecasts (Figure 11a-b). This conclusion holds for a range of forecast horizons up to six weeks ahead and a range of price differentials between peak and off-peak energy prices. This clearly demonstrates that the probabilistic forecasts are sufficiently reliable to improve water management decisions, and is consistent with the identification of statistical forecast skill.

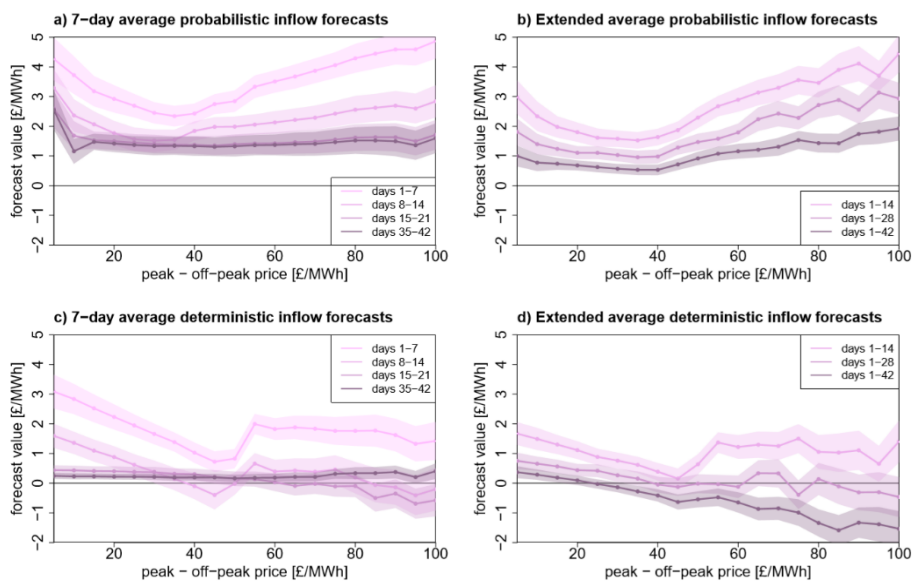

*Figure 11 Economic value of (a-b) Probabilistic and (c-d) deterministic S2S inflow forecasts for different forecast horizons. Plots show the average additional economic value achieved, in terms of unit energy price, relative to deterministic climatological forecasts. The average forecast value is measured over the entire historical observation record (2009-2018). Forecast value is simulated for a range of price differentials between peak and off-peak energy prices, while the peak price is kept constant at £50/MWh. Shaded areas indicate two standard errors (approx. 90% confidence interval), based on bootstrap resampling of data. See supplementary material for further details.*

The highest net unit-rate achieved in the stylised cost model framework is through the application of S2S probabilistic 6-week (days 1-42) average inflow forecasts (not shown), even though these forecasts deliver lower additional economic value compared with forecasts of shorter duration (Figure 11a-b). This is because, we measure the additional economic value delivered by the probabilistic inflow forecasts relative to deterministic climatological forecasts. The average unit rate achieved by the climatological forecasts increases significantly for forecasts of longer duration. For example, a climatological forecast for the average inflow rate over a six week period, will achieve a higher unit rate compared with using climatological forecasts of the average inflow rate over a single week. This is because there is significantly less inter-annual variability in the inflow rate averaged over an entire month, compared with a single week within that month. Thus, a climatological forecast for the entire month of January will be more accurate and result in smaller errors, compared with a climatological forecast for an individual week within January.

The stylised cost model framework reveals that the value of the S2S probabilistic inflow forecasts is non-linearly related to the price differential between peak and off-peak energy prices (Figure 11). Probabilistic forecasts deliver the largest increase in economic value, relative to climatological forecasts, when the price differential is either very high (greater than £60/MWh) or low (less than £20/MWh). This result indicates that probabilistic inflow forecasts may become increasingly valuable if future energy prices are more volatile, as they are expected to in the coming years. When the difference between peak and off-peak energy prices is high, extending power generation into off-peak periods incurs a large opportunity cost. This means there is a high cost for taking precautionary action to keep the reservoir at its target water level, such as extending generation (i.e. discharge) into off-peak periods to accommodate forecast high inflows. In addition, there are substantial costs associated with inaccurate forecast information that result in deviations from the target water level. Under these conditions, probabilistic forecasts are essential to guide optimal water-management decisions. This is because probabilistic forecasts allow operators to fully consider risks posed by high-cost, low-probability outcomes.

The S2S deterministic inflow forecasts consistently deliver lower economic value compared with contemporary probabilistic forecasts (Figure 11). Moreover, under certain conditions the S2S deterministic inflow forecasts deliver negative forecast value relative to the application of climatological forecasts (Figure 11c-d). Thus, the application of S2S deterministic inflow forecasts can lead to poor water management decisions that result in economic losses. This is true even for the 7-

day average deterministic inflow forecast for week 2 (days 8-14). In contrast with probabilistic forecasts, deterministic inflow forecasts hold lower economic value when the price differential is high. This is because deterministic forecasts do not account for the possibility of high-cost, low-probability outcomes, which become increasingly important for higher price differentials. The deterministic inflow forecasts deliver the greatest economic value when the price differential is low (less than £20 / MWh). This is because power sold during off-peak periods will achieve a similar price to power sold during peak periods. For example, generation from a hydropower reservoir can be extended from 12 hours per day to 24 hours per day, to accommodate forecast high inflows, with limited financial implications. Hence, when the price differential is low the cost of taking precautionary action is low. Similarly, there are limited costs associated with inaccurate forecast information that result in deviations from the target water level, unless a spill occurs.

It is important to note that the stylised cost model used here is a highly simplified representation of current operations. An important limitation is that we treat each forecast individually and do not account for possible accumulations of forecast errors over time. In addition, some of the economic value that is highlighted within the S2S probabilistic inflow forecasts is likely already realised in present operations through situational awareness of plant operators and meteorologists.

To summarise, in the stylised cost model framework, the S2S probabilistic inflow forecasts are found to consistently deliver increased economic value relative to climatological and deterministic forecasts, in terms of the mean unit rate energy price achieved for the same water resource. This demonstrates that the S2S probabilistic inflow forecasts are sufficiently reliable to improve water-management decisions. The value probabilistic forecasts is generally found to increase with higher price differentials between peak and off-peak energy prices, while the value of deterministic forecasts decreases under these situations and in some cases is negative. This demonstrates how the asymmetry of costs associated with forecast errors necessitates the use of probabilistic forecast information.

## 5.2 Sources of S2S forecast skill and implications for the hydropower sector

Several recent studies have found higher skill for streamflow and precipitation forecasts in the UK and Europe during winter months, compared with summer (Arnal et al. 2018; Bell et al. 2017; Svensson et al. 2015; Weisheimer and Palmer 2014). Teleconnections between weather patterns in the tropics, such as the Madden Julian Oscillation (MJO) and El Niño-Southern Oscillation (ENSO), have been shown to influence winter weather in the eastern North Atlantic region. These teleconnections occur via linkages with the winter-time polar stratospheric vortex and phase of the NAO, providing sources of predictability on time scales of 10-30 days (Lee et al. 2019). The fact the S2S probabilistic inflow forecasts hold higher skill in winter months, compared with summer, is important for hydropower applications. This is because climatological inflow rates are greatest during winter season, which leads to higher water levels in reservoirs and therefore greater risks of spillages.

The phase of the NAO has a major impact on precipitation across wide areas of Europe, especially in western Scotland during winter (Lavers et al. 2013; Svensson et al. 2015; Hall and Hanna 2018). The NAO has therefore been identified as an important climate index for S2S and seasonal predictions of precipitation and streamflow (Scaife et al. 2014; Svensson et al. 2015; Baker et al. 2018). We find here that winters with a negative phase of the NAO hold the highest skill for our S2S probabilistic inflow forecasts. Negative phases of the NAO index during winter are often associated with colder and drier than average conditions over large parts of the UK. Thus, these periods typically coincide with low inflow and streamflow rates. The winters of 2009-2010 and 2010-2011 are two examples where negative phases of the NAO (Scaife et al. 2014) coincided with prolonged, exceptionally low inflow rates at the case study reservoir. The 4-week average (days 1-28) S2S probabilistic inflow forecasts perform well during the winters of 2009-2010 and 2010-2011, predicting clear signals of below average inflows (Figure 9). The availability of reliable S2S inflow forecasts would likely have greatly assisted the operations of the hydropower sector during these periods, and improved water-management. The reason for long-range predictability during a negative winter NAO may in part be due to the stable and persistent nature of these weather patterns, with low precipitation. Nonetheless, we note that the primary source of skill for the 28-day average inflow forecast is from the first two weeks of the S2S forecasts (Figure 4). There is limited signal of below average inflows from the 7-day average inflow forecasts during week 3 (days 15-21) (Figure 8c).

The skill of inflow forecasts for the case study reservoir is substantially lower during summer months, compared with winter, and limited/no significant skill is available beyond week 2 (days 8-14) (Figure 10). Skill scores during summer are lowest for months with a positive phase of the NAO, and skill scores of inflow forecasts beyond week 2 are significantly below zero (Figure 10c). In Scotland, summer months with a positive NAO index are typically characterised by periods of stable anti-cyclonic conditions, with warm and dry weather. Weisheimer and Palmer (2014) describe the ECMWF seasonal precipitation forecasts as being "dangerously useless" during dry summers in the UK, because they may provide forecast users with inaccurate forecast information. While on the whole dry and warm, these summers may also be interrupted by intense and localised convective precipitation. Due to the smaller-scale processes involved, convective precipitation is more challenging to model and forecast compared with large-scale precipitation. The S2S probabilistic forecasts clearly struggle to predict high summer inflow events, such as summer 2011 (Figure 8). The random initiation of convective storms means these features cannot be forecast reliably beyond a few days ahead. Even for short range forecasts, it is difficult to resolve the exact path of these storms and local topographic interactions, due to the relatively coarse horizontal and vertical resolution of the numerical models used (Anghileri et al. 2019; Scheuerer 2014). There is a greater fraction of convective precipitation in the UK during summer months compared with winter (Lavers et al. 2013). This may also help to explain the lower skill of the S2S probabilistic inflow forecasts during summer. Nonetheless, for hydropower applications, reservoir water levels are typically lower during summer months and therefore there are lower risks of spills. Thus, the costs associated with incorrect forecast information are also lower than in winter.

The existence of skill up to 6-weeks ahead for the S2S probabilistic inflow forecasts is encouraging. Nonetheless, it is important to note that the skill scores of the 7-day average inflow forecasts are low and the prediction intervals are large (Figure 4c and Figure 8c). For forecasts in the S2S range, a sharp (i.e. a confident) forecast distribution is not feasible. Instead, the desired attribute of the S2S forecasts is that the probability distributions are reliable. This is confirmed here by reliability diagrams (Figure 7), and the fact that the S2S probabilistic forecasts consistently deliver added value in our cost-model framework.

## 6. Conclusions

In this study, we present a method of generating S2S probabilistic inflow forecasts for hydropower reservoirs, based on S2S ensemble weather predictions, and apply this method to a case study reservoir in Scotland. The inflow forecasts are produced without the application of a hydrological model. Instead, we train a linear regression model for the observed inflow on to the precipitation rate from the S2S ensemble predictions. We subsequently apply EMOS post-processing techniques that build upon the method of Scheuerer (2014), to produce calibrated probabilistic forecasts of inflow.

S2S probabilistic forecasts for the 7-day mean inflow rate are found to hold fair skill up to 6-weeks ahead (days 35-42), relative to climatological forecasts. While significantly greater than zero, the skill for forecasts weeks 3 to week 6 is relatively low and suggests a marginal improvement upon the skill of climatological inflow distributions. However, greater skill is realised by considering forecasts of the average inflow rate over periods longer than 7-days. For example, probabilistic forecasts of the average inflow rate over the next six weeks (days 1-42) demonstrate good skill relative to climatological forecasts, and hold comparable skill to 7-day average probabilistic inflow forecasts for two weeks ahead (days 8-14).

We develop a stylised cost model to assess the potential economic value of the S2S probabilistic inflow forecasts may deliver to the hydropower sector. This demonstrates that the probabilistic forecasts are sufficiently reliable to consistently improve water management decisions, relative to climatological and deterministic forecasts, delivering a higher average unit-rate for the power generated. In the model framework, the value of the probabilistic forecasts is found to increase for higher price differentials between peak and off-peak energy prices, while the value of deterministic forecasts is lower or even negative under these conditions. This is because probabilistic forecasts allow operators to fully consider the risks of high cost, low probability outcomes. Thus, if energy prices become more volatile, S2S probabilistic inflow forecasts are likely to become increasingly valuable to the hydropower sector.

While the focus of this study is on the hydropower sector in Scotland, the methods outlined here will be relevant for other water-management applications, such as forecasting streamflow in rivers, public water availability, and early warning systems for floods and droughts.

## References

Ahmad, S. K., and F. Hossain, 2019: A generic data-driven technique for forecasting of reservoir inflow: Application for hydropower maximization. *Environ. Model. Softw.*, **119**, 147–165, https://doi.org/10.1016/j.envsoft.2019.06.008.

Alexander, S., G. Yang, G. Addisu, and P. Block, 2021: Forecast-informed reservoir operations to guide hydropower and agriculture allocations in the Blue Nile basin, Ethiopia. *Int. J. Water Resour. Dev.*, **37**, 208–233, https://doi.org/10.1080/07900627.2020.1745159.

Anghileri, D., S. Monhart, C. Zhou, K. Bogner, A. Castelletti, P. Burlando, and M. Zappa, 2019: The Value of Subseasonal Hydrometeorological Forecasts to Hydropower Operations: How Much Does Preprocessing Matter? *Water Resour. Res.*, **55**, 10159–10178, https://doi.org/10.1029/2019WR025280.

Arnal, L., H. L. Cloke, E. Stephens, F. Wetterhall, C. Prudhomme, J. Neumann, B. Krzeminski, and F. Pappenberger, 2018: Skilful seasonal forecasts of streamflow over Europe? *Hydrol. Earth Syst. Sci.*, **22**, 2057–2072, https://doi.org/10.5194/hess-22-2057-2018.

Baker, L. H., L. C. Shaffrey, and A. A. Scaife, 2018: Improved seasonal prediction of UK regional precipitation using atmospheric circulation. *Int. J. Climatol.*, **38**, e437–e453, https://doi.org/10.1002/joc.5382.

Baker, S. A., A. W. Wood, and B. Rajagopalan, 2019: Developing Subseasonal to Seasonal Climate Forecast Products for Hydrology and Water Management. *JAWRA J. Am. Water Resour. Assoc.*, **55**, 1024–1037, https://doi.org/10.1111/1752-1688.12746.

Bazile, R., M.-A. Boucher, L. Perreault, and R. Leconte, 2017: Verification of ECMWF System4 for seasonal hydrological forecasting in a northern climate. *Hydrol. Earth Syst. Sci. Discuss.*, 1–22, https://doi.org/10.5194/hess-2017-387.

Bell, V. A., H. N. Davies, A. L. Kay, A. Brookshaw, and A. A. Scaife, 2017: A national-scale seasonal hydrological forecast system: development and evaluation over Britain. *Hydrol. Earth Syst. Sci.*, **21**, 4681–4691, https://doi.org/10.5194/hess-21-4681-2017.

Branković, Č., T. N. Palmer, F. Molteni, S. Tibaldi, and U. Cubasch, 1990: Extended-range predictions with ECMWF models: Time-lagged ensemble forecasting. *Q. J. R. Meteorol. Soc.*, **116**, 867–912, https://doi.org/10.1002/qj.49711649405.

Brayshaw, D. J., A. Halford, S. Smith, and K. Jensen, 2020: Quantifying the potential for improved management of weather risk using sub-seasonal forecasting: The case of UK telecommunications infrastructure. *Meteorol. Appl.*, **27**, https://doi.org/10.1002/met.1849.

Buizza, R., 2018: Ensemble Forecasting and the Need for Calibration. *Statistical Postprocessing of Ensemble Forecasts*, Elsevier, 15–48.

——, and M. Leutbecher, 2015: The forecast skill horizon. *Q. J. R. Meteorol. Soc.*, **141**, 3366–3382, https://doi.org/10.1002/qj.2619.

Contreras, E., J. Herrero, L. Crochemore, I. Pechlivanidis, C. Photiadou, C. Aguilar, and M. J. Polo, 2020: Advances in the Definition of Needs and Specifications for a Climate Service Tool Aimed at Small Hydropower Plants' Operation and Management. *Energies*, **13**, 1827, https://doi.org/10.3390/en13071827.

Ferro, C. A. T., 2014: Fair scores for ensemble forecasts. *Q. J. R. Meteorol. Soc.*, **140**, 1917–1923, https://doi.org/10.1002/qj.2270.

Gneiting, T., and M. Katzfuss, 2014: Probabilistic Forecasting. *Annu. Rev. Stat. Its Appl.*, **1**, 125–151, https://doi.org/10.1146/annurev-statistics-062713-085831.

Hall, R. J., and E. Hanna, 2018: North Atlantic circulation indices: links with summer and winter UK temperature and precipitation and implications for seasonal forecasting. *Int. J. Climatol.*, **38**, e660–e677, https://doi.org/10.1002/joc.5398.

Harrigan, S., C. Prudhomme, S. Parry, K. Smith, and M. Tanguy, 2018: Benchmarking ensemble streamflow prediction skill in the UK. *Hydrol. Earth Syst. Sci.*, **22**, 2023–2039, https://doi.org/10.5194/hess-22-2023-2018.

Hersbach, H., and Coauthors, 2020: The ERA5 global reanalysis. *Q. J. R. Meteorol. Soc.*, 1–51, https://doi.org/10.1002/qj.3803.

Ho, L. T. T., L. Dubus, M. De Felice, and A. Troccoli, 2020: Reconstruction of Multidecadal Country-Aggregated Hydro Power Generation in Europe Based on a Random Forest Model. *Energies*, **13**, 1786, https://doi.org/10.3390/en13071786.

Huertas-Hernando, D., and Coauthors, 2017: Hydro power flexibility for power systems with variable renewable energy sources: an IEA Task 25 collaboration. *Wiley Interdiscip. Rev. Energy Environ.*, **6**, e220, https://doi.org/10.1002/wene.220.

Jones, P. D., T. Jonsson, and D. Wheeler, 1997: Extension to the North Atlantic oscillation using early instrumental pressure observations from Gibraltar and south-west Iceland. *Int. J. Climatol.*, **17**, 1433–1450, https://doi.org/10.1002/(sici)1097-0088(19971115)17:13<1433::aid-joc203>3.3.co;2-g.

Khouja, M., 1999: The single-period (news-vendor) problem: Literature review and suggestions for future research. *Omega*, **27**, 537–553, https://doi.org/10.1016/S0305-0483(99)00017-1.

Klemm, T., and R. A. McPherson, 2017: The development of seasonal climate forecasting for agricultural producers. *Agric. For. Meteorol.*, **232**, 384–399, https://doi.org/10.1016/j.agrformet.2016.09.005.

Lavers, D., C. Prudhomme, and D. M. Hannah, 2013: European precipitation connections with large-scale mean sea-level pressure (MSLP) fields. *Hydrol. Sci. J.*, **58**, 310–327, https://doi.org/10.1080/02626667.2012.754545.

Lee, R. W., S. J. Woolnough, A. J. Charlton-Perez, and F. Vitart, 2019: ENSO Modulation of MJO Teleconnections to the North Atlantic and Europe. *Geophys. Res. Lett.*, **46**, 13535–13545, https://doi.org/10.1029/2019GL084683.

Lorenz, C., and H. Kunstmann, 2012: The Hydrological Cycle in Three State-of-the-Art Reanalyses: Intercomparison and Performance Analysis. *J. Hydrometeorol.*, **13**, 1397–1420, https://doi.org/10.1175/JHM-D-11-088.1.

Magnusson, J., G. Nævdal, F. Matt, J. F. Burkhart, and A. Winstral, 2020: Improving hydropower inflow forecasts by assimilating snow data. *Hydrol. Res.*, **51**, 226–237, https://doi.org/10.2166/nh.2020.025.

Manrique-Suñén, A., N. Gonzalez-Reviriego, V. Torralba, N. Cortesi, and F. J. Doblas-Reyes, 2020: Choices in the Verification of S2S Forecasts and Their Implications for Climate Services. *Mon. Weather Rev.*, **148**, 3995–4008, https://doi.org/10.1175/MWR-D-20-0067.1.

Merryfield, W. J., and Coauthors, 2020: Current and emerging developments in subseasonal to decadal prediction. *Bull. Am. Meteorol. Soc.*, **101**, E869–E896, https://doi.org/10.1175/BAMS-D-19-0037.1.

Messner, J. W., P. Pinson, J. Browell, M. B. Bjerregård, and I. Schicker, 2020: Evaluation of wind power forecasts—An up-to-date view. *Wind Energy*, **23**, 1461–1481, https://doi.org/10.1002/we.2497.

Ødegård, H. L., J. Eidsvik, and S.-E. Fleten, 2019: Value of information analysis of snow measurements for the scheduling of hydropower production. *Energy Syst.*, **10**, 1–19, https://doi.org/10.1007/s12667-017-0267-3.

Orsolini, Y. J., I. T. Kindem, and N. G. Kvamstø, 2011: On the potential impact of the stratosphere upon seasonal dynamical hindcasts of the North Atlantic Oscillation: A pilot study. *Clim. Dyn.*,

**36**, 579–588, https://doi.org/10.1007/s00382-009-0705-6.

Palmer, T. N., and D. L. T. Anderson, 1994: The prospects for seasonal forecasting—A review paper. *Q. J. R. Meteorol. Soc.*, **120**, 755–793, https://doi.org/10.1002/qj.49712051802.

Rigby, R. A., and D. M. Stasinopoulos, 2005: Generalized additive models for location, scale and shape (with discussion). *J. R. Stat. Soc. Ser. C (Applied Stat.*, **54**, 507–554, https://doi.org/10.1111/j.1467-9876.2005.00510.x.

Scaife, A. A., and Coauthors, 2014: Skillful long-range prediction of European and North American winters. *Geophys. Res. Lett.*, **41**, 2514–2519, https://doi.org/10.1002/2014GL059637.

Scheuerer, M., 2014: Probabilistic quantitative precipitation forecasting using Ensemble Model Output Statistics. *Q. J. R. Meteorol. Soc.*, **140**, 1086–1096, https://doi.org/10.1002/qj.2183.

Sene, K., 2016: *Hydrometeorology*. Springer International Publishing,.

Svensson, C., 2016: Seasonal river flow forecasts for the United Kingdom using persistence and historical analogues. *Hydrol. Sci. J.*, **61**, 19–35, https://doi.org/10.1080/02626667.2014.992788.

Svensson, C., and Coauthors, 2015: Long-range forecasts of UK winter hydrology. *Environ. Res. Lett.*, **10**, https://doi.org/10.1088/1748-9326/10/6/064006.

Vitart, F., 2004: Monthly Forecasting at ECMWF. *Mon. Weather Rev.*, **132**, 2761–2779, https://doi.org/10.1175/MWR2826.1.

——, 2014: Evolution of ECMWF sub-seasonal forecast skill scores. *Q. J. R. Meteorol. Soc.*, **140**, 1889–1899, https://doi.org/10.1002/qj.2256.

——, and A. W. Robertson, 2018: The sub-seasonal to seasonal prediction project (S2S) and the prediction of extreme events. *npj Clim. Atmos. Sci.*, **1**, 3, https://doi.org/10.1038/s41612-018-0013-0.

Vitart, F., and Coauthors, 2017: The subseasonal to seasonal (S2S) prediction project database. *Bull. Am. Meteorol. Soc.*, **98**, 163–173, https://doi.org/10.1175/BAMS-D-16-0017.1.

Wanner, H., S. Brönnimann, C. Casty, D. Gyalistras, J. Luterbacher, C. Schmutz, D. B. Stephenson, and E. and Xoplaki, 2001: NORTH ATLANTIC OSCILLATION – CONCEPTS AND STUDIES. *Surv. Geophys.*, **22**, 321–381, https://doi.org/https://doi.org/10.1023/A:1014217317898.

Weisheimer, A., and T. N. Palmer, 2014: On the reliability of seasonal climate forecasts. *J. R. Soc. Interface*, **11**, https://doi.org/10.1098/rsif.2013.1162.

White, C. J., and Coauthors, 2017: Potential applications of subseasonal-to-seasonal (S2S) predictions. *Meteorol. Appl.*, **24**, 315–325, https://doi.org/10.1002/met.1654.

Wood, A. W., and D. P. Lettenmaier, 2008: An ensemble approach for attribution of hydrologic prediction uncertainty. *Geophys. Res. Lett.*, **35**, L14401, https://doi.org/10.1029/2008GL034648.

Woolnough, S. J., 2019: The Madden-Julian Oscillation. *Sub-Seasonal to Seasonal Prediction*, Elsevier, 93–117.

Yang, T., A. A. Asanjan, E. Welles, X. Gao, S. Sorooshian, and X. Liu, 2017: Developing reservoir monthly inflow forecasts using artificial intelligence and climate phenomenon information. *Water Resour. Res.*, **53**, 2786–2812, https://doi.org/10.1002/2017WR020482.

Zhang, F., Y. Q. Sun, L. Magnusson, R. Buizza, S.-J. Lin, J.-H. Chen, and K. Emanuel, 2019: What Is the Predictability Limit of Midlatitude Weather? *J. Atmos. Sci.*, **76**, 1077–1091,

https://doi.org/10.1175/JAS-D-18-0269.1.

European Centre for Medium-Range Weather Forecasts. (2017, updated monthly) 'ERA5 Reanalysis', Research Data Archive at the National Center for Atmospheric Research, Computational and Information Systems Laboratory. https://doi.org/10.5065/D6X34W69.

# The application of sub-seasonal-to-seasonal (S2S) predictions for hydropower forecasting

Supplementary Material

## DESCRIPTION OF STYLISED COST MODEL

To explore whether the subseasonal-to-seasonal (S2S) inflow forecasts can improve water-management decisions, we develop a stylised cost model, based on the classical *"News Vendor"* optimisation problem (Khouja, 1999). The cost model is guided by current operational practice, with the underlying principle of keeping the reservoir at a target water level (which may vary with seasons).

In the model, an initial cost (stage 1) is incurred for any adjustments to the planned generation schedule (i.e. discharge from the reservoir). These are costs that results from precautionary actions taken based on forecast information. An additional cost (stage 2) is subsequently incurred for any deviation from the target water level at the end of the forecast period (i.e. inflow – discharge). These are costs that result from the realised inflow, following any precautionary actions taken.

The cost model is applied to every forecast within the 10-year observational time-series (2 forecasts per week) from 2009-2019. Based on each inflow forecast, a water-management decision is made on whether to make an adjustment to the planned generation schedule over the forecast period, in order to keep the reservoir at its target water level and maximise the value of water through power generation.

The relative economic value of the different types of forecasts (climatological, deterministic and probabilistic), is evaluated by comparing the total costs incurred for all forecasts within the 10-year time series.

### Cost of adjustments to the generation schedule

Consider an example where a forecast indicates high inflow over the forecast period. Based on this forecast, a decision may be taken to increase generation (i.e. discharge from the reservoir). This would help accommodate the additional inflow and prevent a positive water level anomaly at the end of the forecast period. This adjustment ($A$) to the generation schedule over the forecast period is a form of precautionary action. Such precautionary action may incur a cost ($C_1$) at Stage 1 in the cost model. Details of these costs are presented in **Figure S12** and described below.

The model allows generation to be increased by up to 20% without any costs incurred (**Figure S12**). For example, generation could be increased from 10 hours per day to 12 hours per day, without any financial implications. This is under the assumption that the additional 2 hours of power generated per day may be sold at the peak energy price. However, beyond this 20% threshold, it is assumed that any additional power generated will be sold at the off-peak price (**Figure S12**). Hence, beyond a 20% increase in generation, there is an opportunity cost ($C_1$) associated with the precautionary action ($A$). This cost is equal to the price differential between peak and off-peak energy prices, multiplied by the additional power generated beyond the 20% threshold. Hydropower facilities have a maximum operating capacity (i.e. operating at full capacity 24 hours per day). Any additional discharge beyond this capacity cannot be used to generate power and is considered spill. This spillage is therefore lost revenue costing the peak energy price times the volume of lost water per unit of power (**Figure S12**).

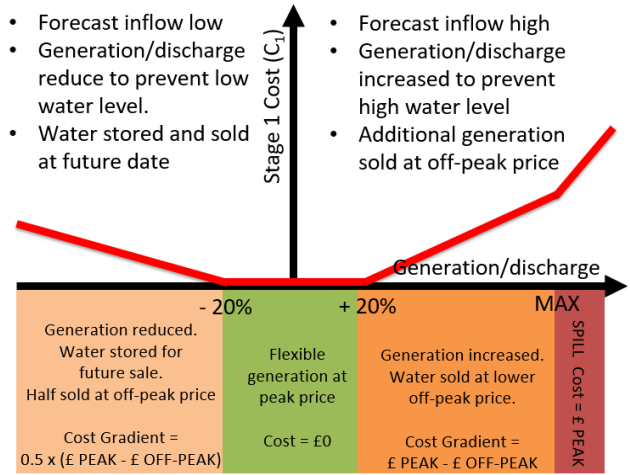

**Figure S12: Cost of precautionary action: adjustments to generation schedule**
*Illustration of stage 1 costs within the model for taking precautionary action to avoid deviations from target water level or spillages. Stage 1 costs are proportional to the price differential between peak and off-peak energy prices, and dependent on the choice of adjustment.*

If a forecast indicates low inflow over the forecast period, a decision may be taken to reduce generation to prevent a negative water level anomaly. In this case, water that would otherwise have been sold now at the peak energy price is stored and will be sold at a future data. Similar to the case above, the cost model allows generation to be reduced by 20% (e.g. from 10 hours per day to 8 hours per day) without incurring any costs. This assumes that the stored water will be sold at a future date at the peak energy price. For larger volumes of water, beyond the 20% threshold, the model assumes that half of the water stored will be sold at off-peak rather than peak energy prices (**Figure S12**). Hence, the cost ($C_1$) of this precautionary action ($A$) is the reduction in generation below 20% multiplied by half the price difference between peak and off-peak energy prices.

## Cost of deviations from the target water level

A high water level limits a reservoirs capacity to impound future high inflows, and increases the risk of spills. Hence, high water levels restrict the flexibility of plant operators. If water levels rise, it becomes increasingly likely that generation will need to be extended for longer periods each day, meaning that more power will be sold at off-peak energy prices; thus, reducing the overall value of water. To account for these factors, additional stage 2 cost ($C_2$) are incurred in the model for deviations from the target water level of the reservoir, at the end of each forecast period (**Figure S13**).

Deviations from the target water level occur when the discharge, following any adjustment to the generation schedule, does not balance the observed inflow. Hence, accurate inflow forecast information and appropriate precautionary action can minimise these costs.

The cost model allows the observed inflow to exceed the adjusted discharge by up to 20% of the climatological discharge rate, without incurring any stage 2 costs. Beyond this 20% threshold, we assume that any additional inflow will be sold at off-peak prices (**Figure S13**). If the observed inflow exceeds the adjusted discharge by more than the maximum operating capacity, we assume that a spill occurs and the volume of water over this threshold is lost with a value of the peak energy price.

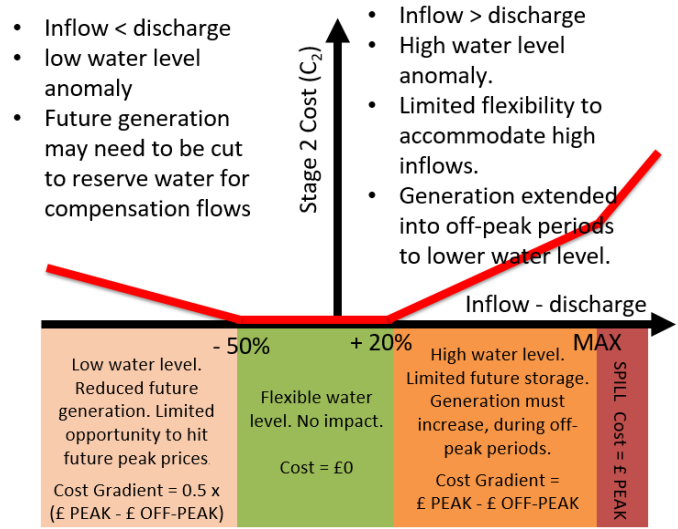

**Figure S13: Cost of deviations from the target water level**
*Illustration of stage 2 costs within the model that are incurred for deviations from the target water level. Stage 2 costs are proportional to the price differential between peak and off-peak energy prices, and dependent on the choice of adjustment and observed inflow. Deviations from the target water level occur if the adjusted generation (discharge) does not balance the observed inflow*

When the reservoir level is low, there is greater capacity to impound future high inflows and to keep generation restricted to peak price periods. Hence, the model assumes greater flexibility when the observed inflow is less than adjusted discharge. In this case, the difference between the inflow and adjusted discharge can reach 50% of the climatological discharge rate without incurring stage 2 costs (**Figure S13**). However, if the reservoir level becomes too low, water may need to be reserved for maintaining compensation flows, leaving no capacity to generate power during peak price periods. If the difference between the observed inflow and adjusted discharge exceeds 50% of the climatological discharge rate, a cost is incurred in the model that is proportional to the volume of water per unit energy multiplied by half the peak and off-peak price difference (**Figure S13**).

## Water-management decisions

For each inflow forecast, a decision must be made on whether to take precautionary action by adjusting ($A$) the planned generation schedule, and if so, by how much (**Figure S14**). The optimal choice of adjustment is that which minimises the combined stage 1 ($C_1(A)$) and expected stage 2 costs ($E\{C_2(A, I_{Forecast})\}$), based on the forecast inflow ($I_{Forecast}$). Hence:

1. $A = \underset{A}{\mathrm{argmin}}(C_1(A) + E\{C_2(A, I_{Forecast})\})$

At the end of each forecast period, the true total cost can be calculated by summing the cost of any precautionary actions (stage 1 costs) and the costs resulting from deviations to the target water level (stage 2 costs), based on the chosen adjustment ($A$) to the generation schedule and the observed inflow ($I_{Observed}$) over the forecast period (Equation 2).

2. $C_{Total} = C_1(A) + C_2(A, I_{Observed})$

To evaluate the economic value of different forecasts, we compare the total costs incurred over the 2009-2019 time series based on the choice of adjustment ($A$) in Equation 1, stemming from the use of: (1) climatological inflow forecasts; (2) deterministic inflow forecasts; (3) probabilistic inflow forecasts.

Climatological and deterministic inflow forecasts predict a single value of inflow for each forecast period. Hence, these forecasts assume a 100% probability that the observed inflow will equal the forecast inflow value. In contrast, probabilistic forecasts predict a range of possible inflows, along with the probability of each inflow occurring. With these probabilistic forecasts, the expected Stage 2 costs in Equation 1 are calculated by weighting each forecast value of inflow by the probability of that inflow value occurring. This method inherently incorporates the risk of possible high cost - low probability outcomes, such as spills, into the choice of the optimal adjustment ($A$). For risk-neutral decisions, precautionary action should only be taken if the probability of the event exceeds the cost of taking precautionary action, divided by the loss that would be incurred if the event occurs and no precautionary action is taken.

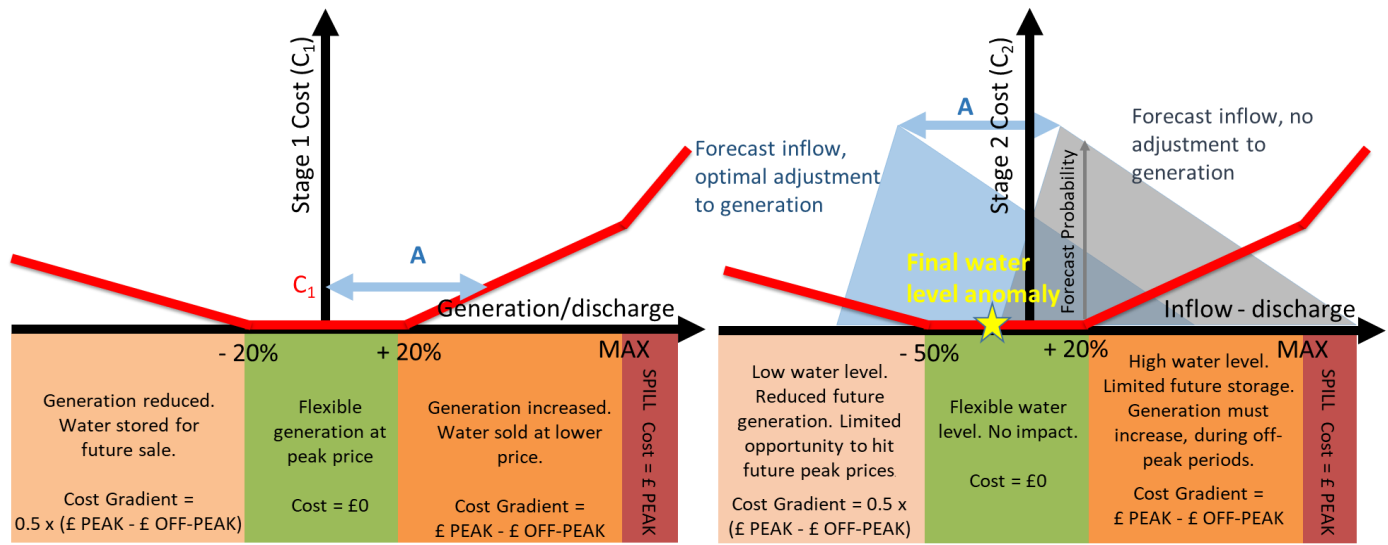

**Figure S14: Choice of optimal adjustment**
*Total cost is equal to the sum of the stage 1 and stage 2 costs. An optimal adjustment (A) to the generation schedule must be determined from each inflow forecast in an attempt to minimise these total costs.*

When comparing the economic value of the three different forecast types, we compare the average unit energy price in £ / MWh achieved over all forecasts in the 10 year time-series. This value is calculated by multiplying the climatological generation by the peak energy price and subtracting the true total cost (Equation 2), before normalising by the climatological generation (Equation 3). The value increase for the probabilistic and deterministic forecasts are measured relative to the use of a climatological forecast.

$$3. \quad Water\ Value\ (£/MWh) = \frac{(GEN_{Clim} \times £\ Peak) - C_{Total}}{GEN_{Clim}}$$

## Experiment set-up

In this model framework, the economic value of the different forecasts is sensitive to the choice of peak and off-peak energy prices (**Figure S14**). To explore this dependency, the cost model is run for a range of price differentials (peak – off-peak) from £5/MWh to £100/MWh, while the peak price is kept constant at £50/MWh. As a result, for differentials over £50/MWh the off-peak price is negative resulting in low "water value" in some cases. In reality, the same differential may also result from a higher peak price and non-negative off-peak price. Therefore, the relative change in water value with price differential is more relevant than the absolute value.

To test the significance of the results, we apply a bootstrap method to resample the 10-year time-series 1000 times (Messner *et al.*, 2020). All results shown indicate the spread of 2 standard errors based on this bootstrap resampling, which approximately represents the 90% confidence interval.

**Performance of climatological forecasts**

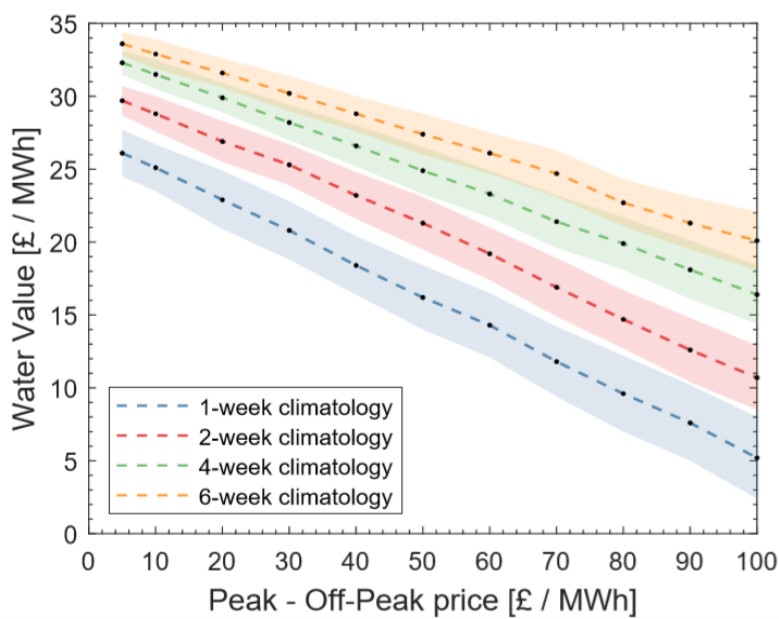

**Figure S15: Average overall value of water achieved for climatological forecasts of different duration**
*Average value of water is shown for a range of price differentials between peak and off-peak energy prices, while the peak price is kept constant at £50/MWh. Shaded areas indicate two standard errors (approx. 90% confidence interval), based on bootstrap resampling of data.*

Here we describe the performance of climatological forecasts in this cost model framework. For all forecast periods, the overall value of water decreases near linearly as the price differential between peak and off-peak energy prices increases (**Figure S15**). This is because the observed inflow is not evenly distributed throughout the year, and the inflow forecasts are not perfect. Therefore, in this model framework, a fraction of water is always sold at off-peak energy prices. It follows that as the off-peak energy price decreases, the overall value of water decreases (**Figure S15**).

In addition to the pattern described above, we see that as the forecast duration increases from one to six weeks, the overall value of the water increases (**Figure S15**). This is for two reasons. The longer duration forecasts have a longer averaging period, meaning that large short-term variations in the observed inflow are smoothed out. Hence, deviations from the climatological mean inflow are reduced, and the performance of climatology as a forecast increases. Furthermore, the longer temporal averaging means that the likelihood of high or low inflow events, requiring precautionary action and/or resulting in deviations from the target water level (i.e. incurring stage 1 or stage 2 costs), is reduced.

Longer duration forecasts (e.g. 6-weeks) are most relevant to reservoirs with a large storage volume. These reservoirs have greater capacity to withstand large short-term variations in inflow, depending on the current water level. For example, a large reservoir would likely have enough storage capacity to impound a period of high inflow over 1-2 weeks, without needing to increase generation into off-peak periods, or enough reserves to continue operating as normal during a period of drought. Hence, these reservoirs require less frequent active water-management.

In contrast, reservoirs with smaller volumes have limited capacity to withstand large short-term variations in inflow. A high inflow event may require generation to be increased from 10 hours per day to 24 hours per day, to accommodate the additional inflow and prevent a spill. This would result in a large fraction of water being sold at off-peak prices. These smaller reservoirs require more active water management, and therefore require detailed additional forecasts information of how the observed inflow is likely to be distributed between the six weeks within a six-week average inflow forecast. Hence, 2-week average inflow forecast are likely to be more relevant for smaller reservoirs.

## REFERENCES

Khouja, M. (1999) 'The single-period (news-vendor) problem: Literature review and suggestions for future research', *Omega*, 27(5), pp. 537–553. doi: 10.1016/S0305-0483(99)00017-1.

Messner, J. W. *et al.* (2020) 'Evaluation of wind power forecasts—An up-to-date view', *Wind Energy*, 23(6), pp. 1461–1481. doi: 10.1002/we.2497.

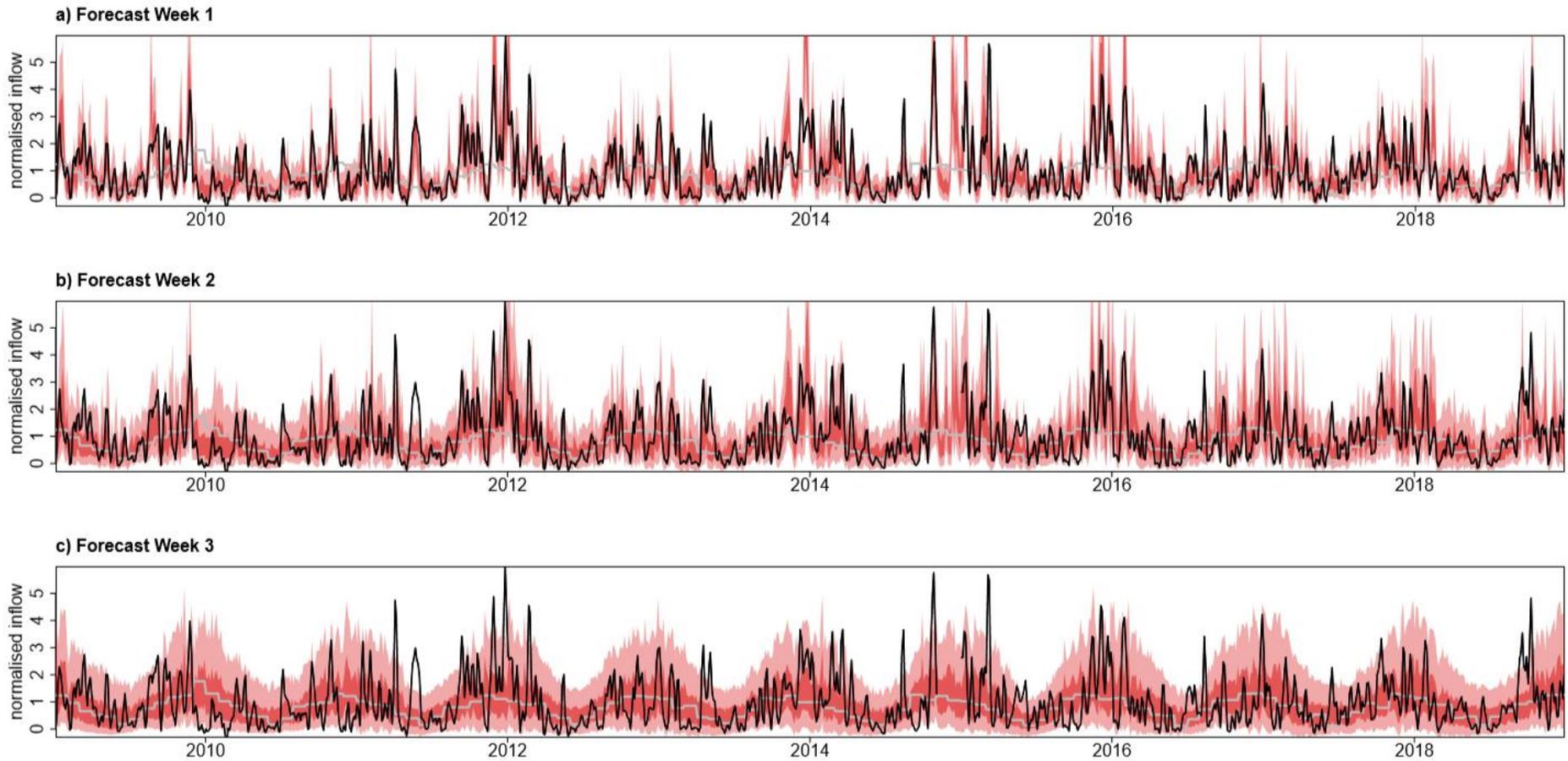

Figure S5 Time series of observed weekly mean inflow rate (black) and the post-processed probabilistic inflow forecast for a) Week 1 (days 1-7), b) Week 2 (days 8-14), and c) Week 3 (days 15-21). Shaded areas indicate the 50% and 90% prediction intervals. Grey line shows the deterministic climatological forecast. Inflow values are normalised by the annual mean inflow.

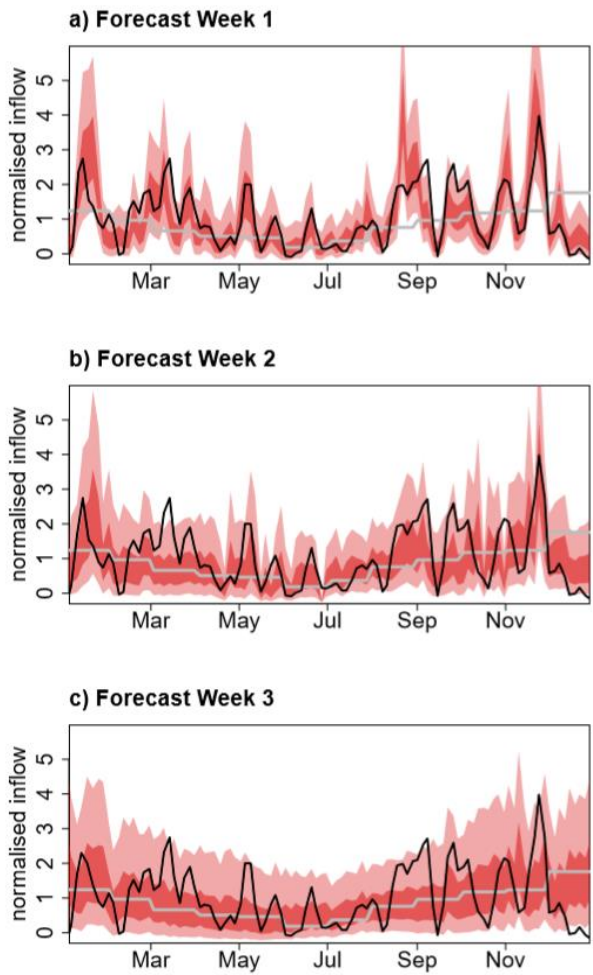

**Figure S6** Time series from 2009 of observed weekly mean inflow rate (black) and the post-processed probabilistic inflow forecast for a) Week 1 (days 1-7), b) Week 2 (days 8-14), and c) Week 3 (days 15-21). Shaded areas indicate the 50% and 90% prediction intervals. Grey line shows the deterministic climatological forecast. Inflow values are normalised by the annual mean inflow.

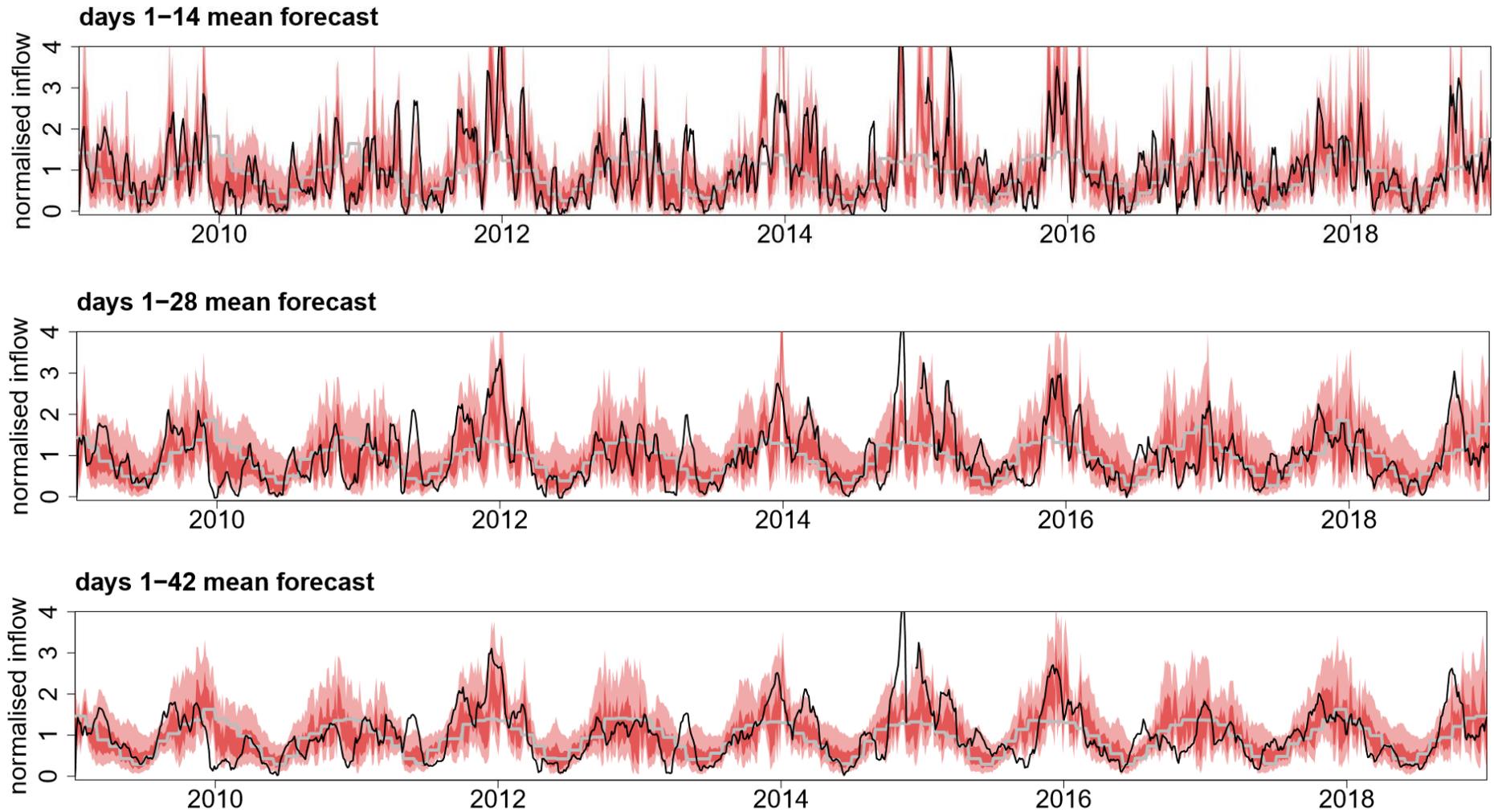

**Figure S7** Time series of observed mean inflow rate (black) and the post-processed probabilistic inflow forecasts a) 2-week forecast (days 1-14), b) 4-week forecast (days 1-28), and c) 6-week forecast (days 1-42). Shaded areas indicate the 50% and 90% prediction intervals. Grey line shows the deterministic climatological forecast. Inflow values are normalised by the annual mean inflow.

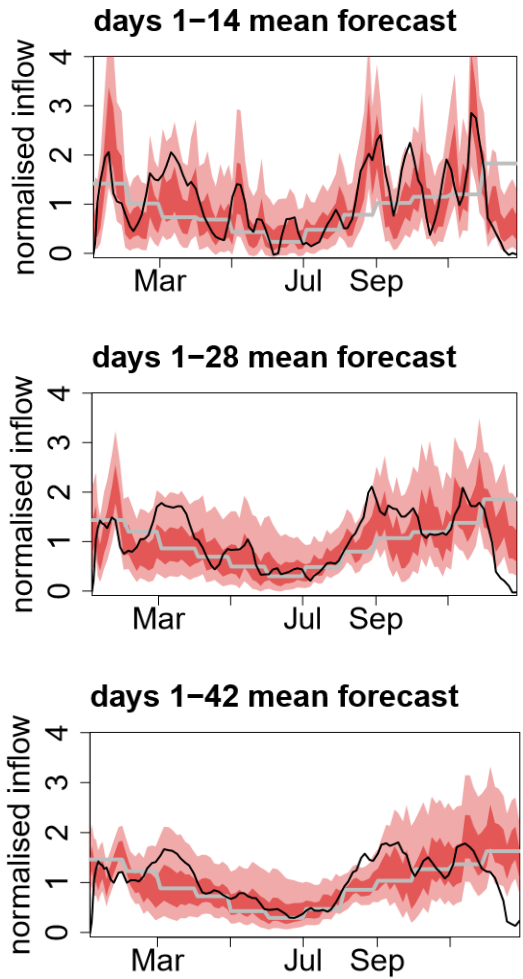

**Figure S8** Time series from 2019 of observed mean inflow rate (black) and the post-processed probabilistic inflow forecasts a) 2-week forecast (days 1-14), b) 4-week forecast (days 1-28), and c) 6-week forecast (days 1-42). Shaded areas indicate the 50% and 90% prediction intervals. Grey line shows the deterministic climatological forecast. Inflow values are normalised by the annual mean inflow.